\documentclass[aps,pra,superscriptaddress,notitlepage,twocolumn,longbibliography]{revtex4-1}
\usepackage[utf8x]{inputenc}
\usepackage[usenames,dvipsnames]{xcolor}
\usepackage{amsmath}
\usepackage{amssymb}
\usepackage{mathtools}
\usepackage{esint}
\usepackage{dcolumn}
\usepackage{bm}
\usepackage{bbm}
\usepackage{blindtext}
\usepackage{braket}
\usepackage{natbib}
\usepackage{graphicx}
\usepackage[normalem]{ulem}
\usepackage{enumerate}
\usepackage{listings}

\usepackage[hidelinks=true,pdfborder={0 0 0},colorlinks=true,linkcolor=blue,urlcolor=blue,citecolor=blue]{hyperref}

\newcommand{\be}{\begin{eqnarray}}
\newcommand{\ee}{\end{eqnarray}}

\newcommand{\veck}{\mathbf k}

\newcommand{\vecq}{\mathbf q}

\newcommand{\vecr}{\mathbf r}
\newcommand{\etal}{\textit{et al.}~}

\newcommand{\ded}{\hat d^\dagger}
\newcommand{\de}{\hat d}
\newcommand{\ced}{\hat c^\dagger}
\newcommand{\ce}{\hat c}

\newcommand{\eqw}[1]{(\ref{#1})}

\newcommand{\fig}[1]{Fig.\thinspace{}\ref{#1}}

\begin{document}

\title{Atomtronics with a spin: statistics of spin transport and non-equilibrium orthogonality catastrophe in cold quantum gases}

\author{Jhih-Shih You}
\affiliation{Department of Physics, Harvard University, Cambridge MA 02138, USA}
\affiliation{Institute for Theoretical Solid State Physics, IFW Dresden, Helmholtzstr. 20, 01069 Dresden, Germany}

\author{Richard Schmidt}
\affiliation{Max Planck Institute of Quantum Optics,  Hans-Kopfermann-Str. 1, 85748 Garching, Germany}

\author{Dmitri A. Ivanov}
\affiliation{Institute for Theoretical Physics, ETH Z\"urich, 8093 Z\"urich, Switzerland}

\author{Michael Knap}
\affiliation{Department of Physics and Institute for Advanced Study, Technical University of Munich, 85748 Garching, Germany}

\author{Eugene Demler}
\affiliation{Department of Physics, Harvard University, Cambridge MA 02138, USA}

\date{\today}
\begin{abstract}
We propose to investigate the full counting statistics of nonequilibrium spin transport with an ultracold atomic quantum gas. The setup makes use of the  spin control available in atomic systems to generate spin transport induced by an impurity atom immersed in a spin-imbalanced two-component Fermi gas. In contrast to solid-state realizations, in ultracold atoms spin relaxation and the decoherence from external sources is largely suppressed. As a consequence, once the spin current is turned off by manipulating the internal spin degrees of freedom of the Fermi system, the nonequilibrium spin population remains constant. Thus one can directly count the number of spins in each reservoir to investigate the full counting statistics of spin flips, which is notoriously challenging in solid state devices. Moreover, using Ramsey interferometry, the dynamical impurity response can be measured. Since the impurity interacts with a many-body environment that is out of equilibrium, our setup provides a  way to realize the non-equilibrium orthogonality catastrophe. Here, even for spin reservoirs initially prepared  in a zero-temperature state, the Ramsey response exhibits an exponential decay, which is in contrast to the conventional power-law decay of Anderson's orthogonality catastrophe. By mapping our system to a multi-step Fermi sea, we are able to derive analytical expressions for the impurity response at late times. This allows us to reveal an intimate connection of the decay rate of the Ramsey contrast and the full counting statistics of spin flips.
\end{abstract}
\maketitle

\section{Introduction}

Some of the most interesting applications of condensed matter theory are concerned with transport~\cite{Beenakker1991,uti2004,Nazarov2009}.  Most  studies of transport focus on averaged quantities such as currents of charge, concentrations, or heat. However, transport experiments contain more information than just those average quantities. Indeed, one of the important ideas that emerged in the studies of transport in condensed matter physics is that fluctuations contain more information than accessible from sole measurements of averaged quantities. In particular, the study of quantum noise that arises from  fluctuations which persist even at zero temperature became of  great practical relevance since it presents the ultimate limit to noise in electronic and spintronic devices. From a more fundamental perspective, the analysis of noise in transport \cite{Blanter2000,Ensslin2006} made the demonstration of charge fractionalization in quantum Hall systems possible \cite{Saminadayar1997,dePicciotto1997}, and provided a new means to separate ballistic and diffusive quasiparticle transport in low-dimensional materials \cite{Agarwal2017}.

Likewise, achieving a high level of control over transport requires a study beyond average quantities \cite{Beenakker2003}. In particular, gaining control on the level of single electrons and spins necessitates the understanding of the intrinsic quantum noise in such systems \cite{Keeling2006, Feve2007, Bennakker2015}. A theoretical tool for this purpose is the full counting statistics~(FCS)  that contains the information about all moments of the desired  observable~\cite{Levitov1996}. In  solid state experiments, the  control of the quantum noise  is, however,  challenging since it is  difficult to change system parameters~\cite{Reulet2003, Bomze2005, Fujisawa2006, Gustavsson2006, Timofeev2007, Gershon2008, Flindt2009, Gabelli2009, Gustavsson2009, Ubbelohde2012}.

In recent years, ultracold atoms have emerged as a toolbox to study the transport of in- and out-of-equilibrium systems in a controlled setting, where a high degree of isolation from the environment  is realized and single-atom resolution is achievable. First examples range from the expansion of fermions in optical lattices \cite{Schneider2012} to the conductivity of a Fermi gas~\cite{anderson_optical_2017}, transport in spin systems~\cite{hild_far--equilibrium_2014, brown_2d_2014} localization induced by disorder in the Hubbard model \cite{Schreiber2015, Kondov2015, bordia_probing_2017}, atom transport in analogs of point contacts \cite{Krinner2014,Husmann2015,Krinner2016} as well as the study of anomalous transport in quantum Hall systems \cite{Li2016,Jotzu2014,Lohse2017}, and topological charge pumping  in bosonic quantum gases \cite{Lohse2015,Schweizer2016}.

Inspired by the recent ultracold atom experiments on quantum impurities~\cite{Schirotzek2009, Nascimbne2009, Kohstall2012, Koschorreck2012, Catani2012, Cetina2015, cetina_ultrafast_2016,Scazza2017, Meinert17}, in this work we propose a new type of transport experiments that allows one to study an analog of spintronics in ultracold atomic systems. Our setup provides a new platform for studying the full counting statistics of transport, and allows one to reveal its remarkable relation to the non-equilibrium orthogonality catastrophe. In contrast to solid state systems, our proposed ultracold atom setup does not suffer from limited coherence times  resulting from phonon relaxation and electron interactions and has the advantage that dynamics takes place on a much longer time scale due to the diluteness of the atomic quantum gas.

\begin{figure}[t]
\centering
\includegraphics[width=\linewidth]{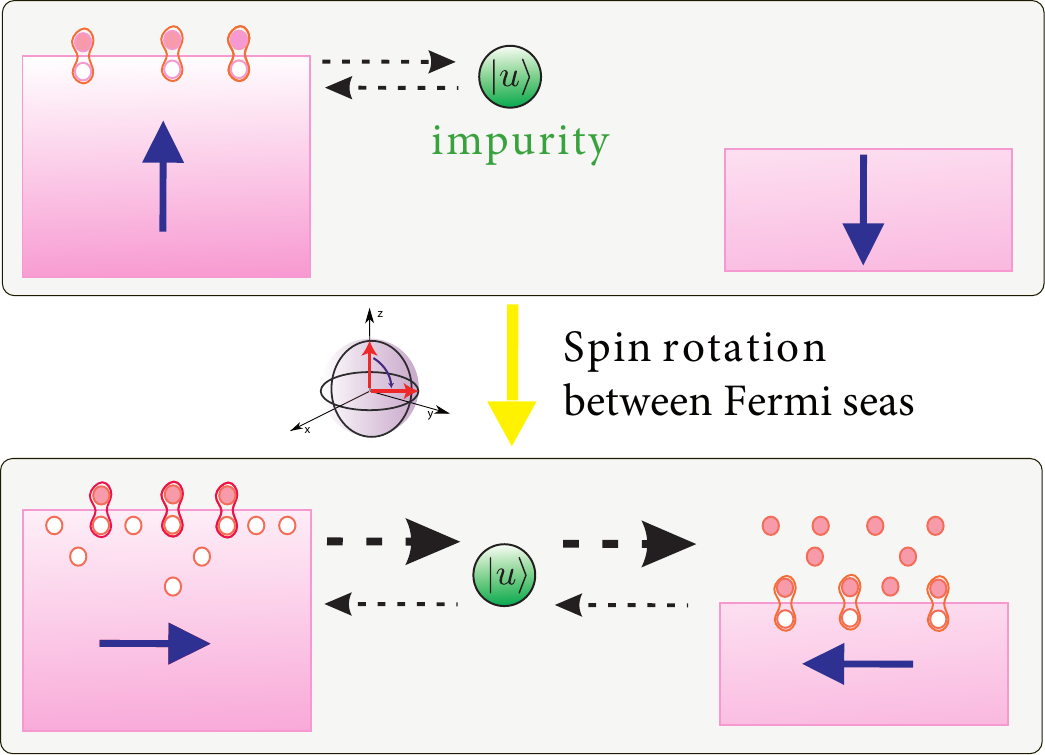}
\caption{\textbf{Schematic representation of our setup.} An
impurity atom is coupled to a spin-imbalanced two-component Fermi gas with mismatched chemical potentials $\mu_{\uparrow} \neq \mu_{\downarrow}.$  Upper row: the impurity atom in the internal state $\ket{u}$ resonantly interacts with the first component $|\uparrow\rangle,$ but not with the second one $|\downarrow\rangle.$  Lower row: applying a spin rotation mixes the two spin states and introduces impurity-induced spin-flips  between the two fermionic components.  
}\label{fig_1}
\end{figure}

Specifically, our proposed setup consists of a single quantum impurity that is coupled to two reservoirs of fermions, see \fig{fig_1}. These two imbalanced Fermi reservoirs can be experimentally realized by preparing fermionic atoms in two different hyperfine states. We show that, despite atom collisions being originally spin conserving, by creating a superposition of the two hyperfine states, spin changing collisions can be engineered. Combined with controllably switching the interactions between the impurity and the Fermi seas, a non-equilibrium spin-flip dynamics between the reservoirs is induced that can be directly measured using, e.g., absorption imaging. Moreover, the full counting statistics of the scattered fermionic particles is accessible, which is characterized by the probability distribution $P_N(t)$ of finding $N$ scattered particles at time $t$. With cold quantum gases this can be achieved  using time-of-flight measurements \cite{bloch_many-body_2008} or quantum gas microscopy \cite{Sherson2010,bakr_quantum_2009,Haller2015,Edge2015,Parsons2016,Boll2016,Cheuk2016,Mazurenko2017}, both techniques that are not available in solid-state systems.

In addition, decoherence dynamics of the system can be studied by applying a Ramsey sequence on the impurity \cite{Knap2012, Cetina2015,Shchadilova2016,Schmidt2017,Ashida2018}. We find that the Ramsey response of the impurity, $S(t)$, is governed by a non-equilibrium orthogonality catastrophe (NOC).  Quite counterintuitively the NOC features an exponential decay in $S(t)$ even at zero temperature. This is in contrast to the conventional orthogonality catastrophe where an exponential decay is a signature of thermal decoherence~\cite{Knap2012} (for a review see Ref.~\cite{Schmidt2017}). Remarkably, in the long-time limit we find, up to logarithmic corrections, a simple relation between the  decay  of the Ramsey signal $S(t)$ and the FCS of spin flips at zero temperature
\begin{equation}
|S(t)|\sim \sqrt{P_{N=0}(t)}. 
\label{eq:relationSP}
\end{equation}
This equation highlights the intimate relation between Ramsey interferometry and the counting statistics of spin flips. \\

This work is organized as follows: In Sec.~\ref{Sec:II} we introduce the model. In Sec.~\ref{Sec:III} we discuss spin transport and full counting statistics for various parameter regimes. 
In Sec.~\ref{Sec:V} we present the results for the impurity decoherence dynamics, which can be measured by Ramsey interferometry and discuss the NOC. The full-time Ramsey response is evaluated numerically, but also long-time analytical expressions are provided. We present an analysis for both zero and finite temperature and  establish the relation between $S(t)$ and $P_{N=0}(t)$. In Sec.~\ref{Sec:VI}  we summarize our results and discuss future prospects.

\section{Model}\label{Sec:II}

We consider a single immobile impurity immersed in a non-interacting two-component Fermi gas. Experimentally the fermionic atoms of mass $m$ are initially prepared in two (hyperfine) spin states  denoted by $(\uparrow,\downarrow)$. Furthermore, we assume that the impurity has two internal states $\ket{u}$ and $\ket{d}$. For simplicity we assume that interactions occur only between the impurity in the state $\ket{u}$ and fermions in the $\ket{\uparrow}$-state; our analysis can, however, be easily generalized. The  Hamiltonian is given by
\begin{align}\label{Hamiltonian1}
\hat H&=&\sum_{\veck\sigma}(\epsilon_\veck-\mu_\sigma) \ced_{\veck\sigma}\ce_{\veck\sigma}
+ \ket{u} \bra{u}\otimes \frac{1}{\mathcal V}\sum_{\veck\vecq} V_{\vecq} \ced_{\veck+\vecq\uparrow} \ce_{\veck\uparrow}
\end{align}
where $\mathcal V$ is the system volume and $\ced_{\veck\sigma}$ and $\ce_{\veck\sigma}$ denote the fermion creation and annihilation operators, respectively. The dispersion relation of the fermions is $\epsilon_\veck=\veck^2/2m$ and their occupation number in the two spin states $\sigma=(\uparrow,\downarrow)$ can be tuned individually by the chemical potentials $\mu_\sigma$.  Unless indicated otherwise we work in units where $\hbar=1$. The short-range potential $V_\vecq$ gives rise to an s-wave scattering phase shift $\delta_k$ for low scattering momenta $k=|\veck|$. While our analytical results hold for general  $\delta_k$, for our numerical results we consider 
\begin{equation}
\delta_k=-\tan^{-1}(ak)
\end{equation}
with the  s-wave scattering length $a$.

\begin{figure*}[t]
	\centering
	\includegraphics[width=0.96\linewidth]{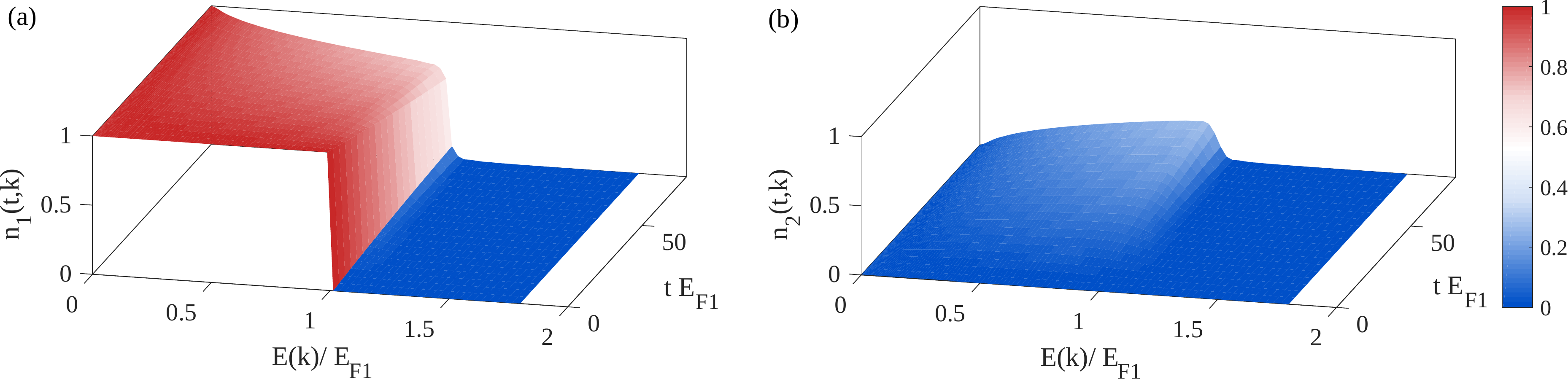}
\caption{ \textbf{Non-equilibrium momentum population.} Occupations (a) $n_1(t,k)$ and (b) $n_2(t,k)$ in the two different fermionic spin states at zero temperature.  Initially, only the state $|1\rangle$ is occupied up to the Fermi energy and no atom are in the second state $|2\rangle.$  We have chosen the dimensionless interaction parameter $k_{F1} a=-6$.}\label{fig_2}
\end{figure*}

The Hamiltonian \eqw{Hamiltonian1} conserves spin and hence does not suffice to study spin transport. In order to introduce the required spin-changing interactions we make use of coherent spin-control available in atomic systems.  To this end we start from the state  $\ket{\text{FS}_\uparrow}\otimes\ket{\text{FS}_\downarrow}$, where  $\ket{\text{FS}_\sigma}$ represent filled Fermi seas (at zero temperature). Then a spin rotation is applied that rotates the spin state of fermions on the Bloch sphere at an arbitrary ``polarization angle'' $\theta$  leading to atoms in a superposition state described by
\begin{align}
\de_{\veck1}&=\cos(\theta/2)\ce_{\veck\uparrow}-\sin(\theta/2)\ce_{\veck\downarrow}\nonumber,\\
 \de_{\veck2}&=\sin(\theta/2)\ce_{\veck\uparrow}+\cos(\theta/2)\ce_{\veck\downarrow}
\end{align}
(for an illustration see Fig~\ref{fig_1}). In the absence of impurities in the $\ket{u}$ state this process is fully coherent. It initializes  the  state $\ket{\psi_F} \equiv \ket{\text{FS}_1}\otimes  \ket{\text{FS}_2}$ with $\ket{\text{FS}_\alpha}=\prod_{|\veck|<k_{F\alpha}} \ded_{\veck \alpha}\ket{0}$ and $\alpha=(1,2)$,
where the Fermi momenta $k_{F1}=k_{F\uparrow}$ and $k_{F2}=k_{F\downarrow}$ are invariant under the spin rotation (similarly, $\mu_1=\mu_\uparrow$ and $\mu_2=\mu_\downarrow$).

Expressing the fermionic operators in Eq.~\eqref{Hamiltonian1} in terms of $\de_{\veck1}$ and $\de_{\veck2}$ yields
\begin{widetext}
\begin{equation}\label{SpinPumpHamiltonian}
\hat H =\underbrace{\overbrace{\sum_{\veck\alpha} \epsilon_{\veck}\ded_{\veck\alpha} \de_{\veck\alpha}}^{\hat H_0}
+\frac{1}{\mathcal V}\sum_{\veck\vecq}  V_{\vecq}\begin{pmatrix}\ded_{\veck+\vecq1} \\ \ded_{\veck+\vecq2}\end{pmatrix}^T \begin{pmatrix}\cos^2(\frac{\theta}{2}) & \cos(\frac{\theta}{2})\sin(\frac{\theta}{2}) \\
 \cos(\frac{\theta}{2})\sin(\frac{\theta}{2})  & \sin^2(\frac{\theta}{2})
 \end{pmatrix}
 \begin{pmatrix}\de_{\veck1} \\ \de_{\veck2}\end{pmatrix}}_{\hat H_1}\otimes\ket{u}\bra{u}
 -\sum_{\alpha} \mu_{\alpha,\veck} \ded_{\veck\alpha}\de_{\veck\alpha}.
 \end{equation}
\end{widetext}
Here, the second term generates spin flip processes between the states ${1}$ to ${2}$ of the atoms in the Fermi seas when scattering with the impurity and thus Eq.~\eqref{SpinPumpHamiltonian} allows one to realize the analog of a  quantum spin pump. In App.~\ref{sec:appendixA} we provide a solution to the single-particle problem corresponding to the Hamiltonian \eqw{SpinPumpHamiltonian} where the spin-dependent interaction is controlled by the polarization angle $\theta$ and interaction strength $V_0$. Both are fully tunable in real time in ultracold atomic systems. In the following we study the dynamical and statistical properties of this Hamiltonian.

\section{Spin transport}\label{Sec:III}

In our setup the Fermi seas $\ket{\text{FS}_1}$ and $\ket{\text{FS}_2}$ represent two `spin reservoirs' 1 and 2. We choose $E_{F2}=0$ so that the system is initially far from the state of equal spin population. Switching the impurity state from  $\ket{d}$ to $\ket{u}$ leads to spin flips that  result in a `spin current'  from reservoir 1 to 2.

\subsection{Spin current}

First we study the `discharging' dynamics of the two-component Fermi gas. In our setup the spin transport rate (we denote it as `spin current') between the reservoirs $\ket{\text{FS}_1}$ and $\ket{\text{FS}_2}$ is controlled by  the rotation angle $\theta$. There are two  processes contributing to the dynamics: First, a fermion in reservoir 1 can scatter with the impurity leading to a change in its momentum state, while it remains in the same spin state. This is a spin-conserving process. By contrast, in the second type of process the impurity can additionally flip its spin in the scattering event, leading to a transfer of spins from reservoir 1 to 2. 

In the time evolution, the spin current generated by the spin flips is accompanied by a build-up of a non-trivial momentum distribution in both spin components $n_{1,2}(\veck,t)=\langle \psi_F|e^{i \hat{H} t}\hat{n}_{1,2}(\veck) e^{-i \hat{H} t}|\psi_F\rangle$. We consider an ultracold, dilute Fermi gas and short-range interactions. Hence only s-wave states contribute to the dynamics and we will only consider these modes in the following. The two main processes contributing to the dynamics are reflected in the s-wave contributions $n_{1,2}(t,k)$ shown in Fig.~\ref{fig_2} ($k$ refers to the s-wave radial momentum).   First, in $\ket{\text{FS}_1}$ the sudden switch on of the impurity leads to the generation of particle-hole fluctuations within the Fermi sea that are the origin of the Anderson orthogonality catastrophe~\cite{Anderson1967}. This dynamics which originates from the momentum changing collisions of the fermions with the impurity is well-studied \cite{Knap2012,Schmidt2017}. There is, however, also the second process corresponding to the spin flips between the states ${1}$ to ${2}$, and, since we have chosen the second Fermi sea $\ket{\text{FS}_2}$ to be initially empty, one can attribute all atoms found in the state $2$ to such spin-flip processes.

The spin-flip probability $\Gamma(E)$ inherits its energy dependence from the phase shift $\delta(E)\equiv\delta_{k=\sqrt{2m E}}$ that increases monotonically with energy $E = \veck^2/2m$. It is determined by recognizing that scattering occurs according to $\ket{\uparrow}\otimes\ket{u}\to e^{2 i \delta(E)}\ket{\uparrow}\otimes\ket{u}$, and $\ket{\downarrow}\otimes\ket{u}\to \ket{\downarrow}\otimes\ket{u}$. From this relation it follows (see App.~\ref{app:SpinFlipRate})
 \begin{align}\label{SpinFlipRate}
 \Gamma(E) = \sin^2\theta \sin^2 \delta(E).
 \end{align}
Since the phase shift $\delta(E)$ increases monotonously in magnitude with energy, the spin-flip probability is largest for fermions close to the Fermi surface. Hence we find the largest build up of occupations in the reservoir 2 close to the Fermi energy  $E_{F1}$ of the first Fermi sea.

Experimentally the momentum occupation $n_2(t,k)$ can be measured by transferring the impurity back to its non-interacting state $\ket{d}$ at time $t$  and simultaneously rotating the Fermi seas back to their ${\uparrow,\downarrow}$-states. Following the separation of the spin states $\uparrow$ and $\downarrow$ by a Stern-Gerlach procedure,  the momentum distribution is obtained from a time-of-flight measurement. Since the dynamics has been initialized with an empty reservoir 2, all observed atoms in the atomic ${\downarrow}$-state can be attributed to the spin-flip dynamics. This  allows one to achieve measurements with a high signal-to-noise ratio.

\begin{figure}[t]
	\centering
	\includegraphics[width=0.48\textwidth]{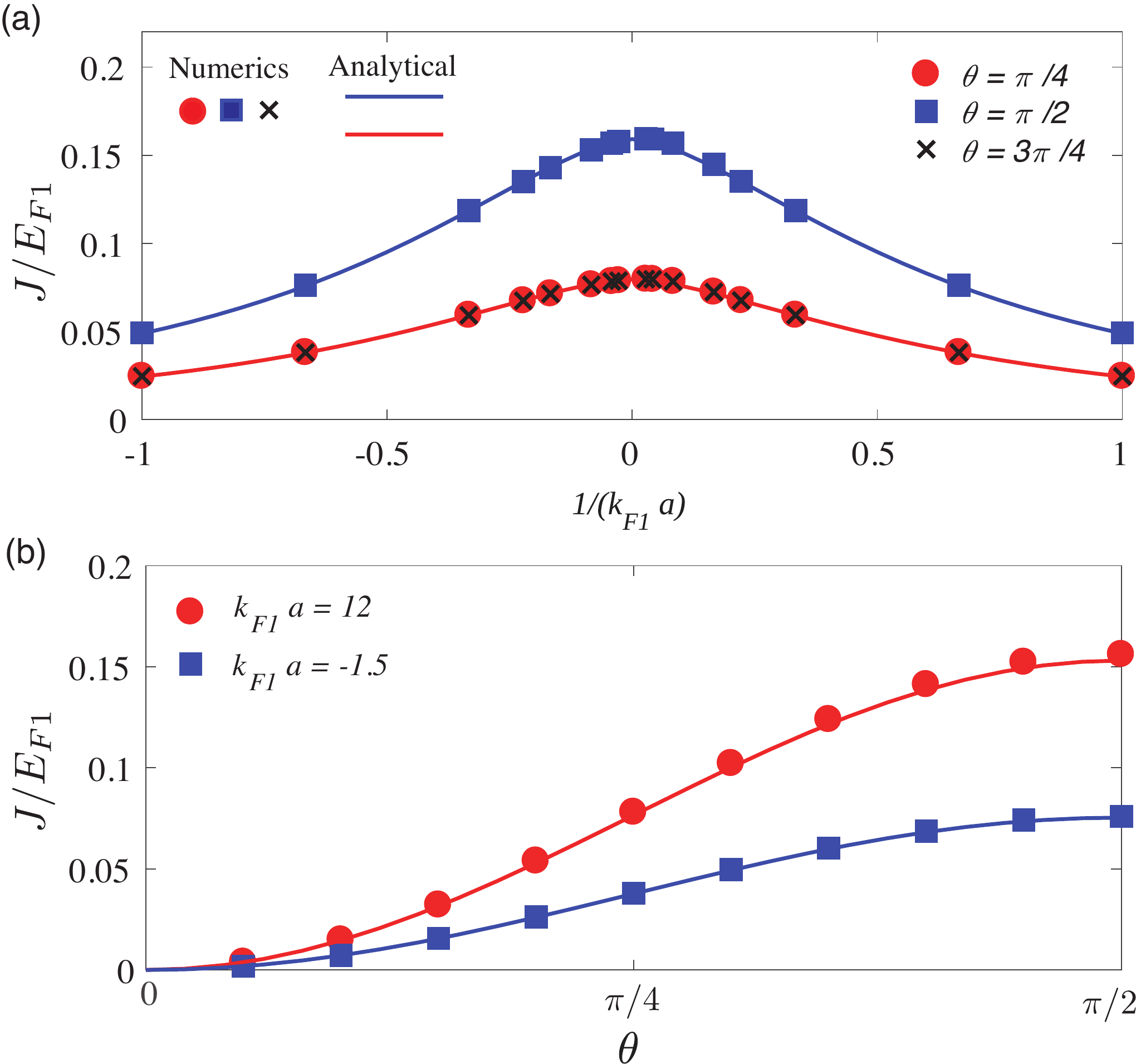}
\caption{\textbf{Non-equilibrium spin current $J$.} (a) The current $J$ is shown for $E_{F2}/E_{F1}=0$ as function of the inverse
scattering length $1/k_{F1} a$ for $\theta=\pi/4, \pi/2, 3\pi/4$. (b) $J$ as function  of the spin rotation $\theta$ for fixed interaction  $k_{F1}a=-1.5$ and $12$. The current $J$ is symmetric with respect to $\theta$ and $\pi-\theta$. Both cases are evaluated at zero temperature $T=0$. The numerically evaluated current $J$, symbols, agrees well with the analytical expression \eqw{anaJ}, solid lines. 
}\label{fig_3}
\end{figure}

We find that the current flow is not only unidirectional from reservoir 1 to 2 at early times, but remains so also at long times. This effect can be understood in a picture where the Fermi sea is decomposed into wave packets that are localized both in energy and space \cite{Schmidt2017}. When these wave packets are scattered off the impurity they move ballistically outwards and can not re-scatter. If their  spin has been flipped in the scattering process they are thus forced to remain in the final spin state. Note that in the scattering process the wave packet becomes a superposition of a spin-flipped and spin-conserved components. In real space this effect will be visible as an ever growing cloud of atoms with spin-flipped components moving outwards from the impurity center.

Summing over the occupation numbers $N_\sigma(t)= \sum_k n_\sigma(k,t)$ we numerically find that after a short initial time a steady current $ N_2(t)= J t$ is established. Here the current is defined as $J=\frac{d\Delta N(t)}{dt}$ with  $\Delta N(t)=N_2(t)-N_2(0)$. The current can also be determined analytically by integrating the spin-flip probability $\Gamma(E)$ in Eq.~\eqref{SpinFlipRate} over the occupation of the reservoir 1. With the phase shift $\delta(E)=-\tan^{-1}(a \sqrt{2 m E}),$ we arrive at
\begin{align}\label{anaJ}
J= \int^{E_{F1}}_0 \frac{dE}{2\pi}\Gamma(E)=\sin^2 \theta\frac{2 m a^2 E_{F1} -\ln(1 + 2 m a^2 E_{F1} )}{4 \pi m a^2}
\end{align}
Fig.~\ref{fig_3} demonstrates that the data, obtained by the functional determinant approach (FDA), see App.~\ref{sec:appendixA} for details, is fully described by the analytical expression. This figure also illustrates how the spin current $J$ can be controlled in various ways.  For instance,  changing the dimensionless scattering length $k_Fa$, the largest current is achieved at resonance where $a$ diverges and the scattering rate is thus maximal. The symmetry between positive and negative values of $k_F a$, directly apparent from the analytical result Eq.~\eqref{anaJ} (cf. also Fig.~\ref{fig_3}(a)), indicates that the bound state, existing for $a>0$ is not relevant for the spin transport dynamics at long times.  Moreover, as shown in Fig.~\ref{fig_3}(b),  the spin current $J$ can  be adjusted by the  polarization angle $\theta$, which determines the ratio of the off-diagonal to diagonal matrix elements in Eq.~\eqref{SpinPumpHamiltonian}. As can be seen from Fig.~\ref{fig_3}(b)  $J$ increases monotonically with $\theta$ and reaches its maximum at $\theta=\pi/2$.

\subsection{Full counting statistics of spin current}\label{Sec:IV}

In solid-state systems it is  notoriously difficult to microscopically observe spin transport dynamics on the level of a few spins. By contrast, with cold atoms one can  directly count the number of transferred spins by absorption imaging. Moreover, spin counting can be achieved in real-time 
by destructively measuring the particle number at arbitrary times because of the characteristically slow dynamics of cold atomic system \cite{cetina_ultrafast_2016}. This brings about the possibility to study time-resolved shot-to-shot fluctuations. 

While the current $J$ gives the \textit{averaged}  particle number $N_2(t)$ transferred per time between the Fermi seas, in each individual experimental measurement the observed number $N_2$ will fluctuate. The corresponding probability  $P_{N_2}$ to measure a certain transferred particle number $N_2$ in an individual experimental realization --- also called the `full counting statistics' (FCS) of $N_2$ --- is given as the Fourier transformation of the characteristic function
\be\label{expansion}
\chi (\lambda,t)&\equiv&\langle e^{i \lambda  \hat{N}}\rangle(t)= \sum_{ N} P_{ N}(t) e^{i \lambda  N}
\ee
with respect to the counting parameter $\lambda$.  

The characteristic function $\chi (\lambda,t)$ contains all information about the distribution of counted particles. In particular arbitrary moments of the distribution $P_N(t)$ can be computed by differentiation $\langle  \hat{N}^n\rangle_t=\frac{d^n}{d (i \lambda )^n }  \chi (\lambda,t)|_{\lambda=0}$. Since $\hat{N}$ is a bilinear, one can compute $\chi (\lambda,t)$ exactly using the functional determinant approach (FDA), from which $P_N(t)$ then follows from a Fourier transform.

\begin{figure}
	\centering
	\includegraphics[width=0.48\textwidth]{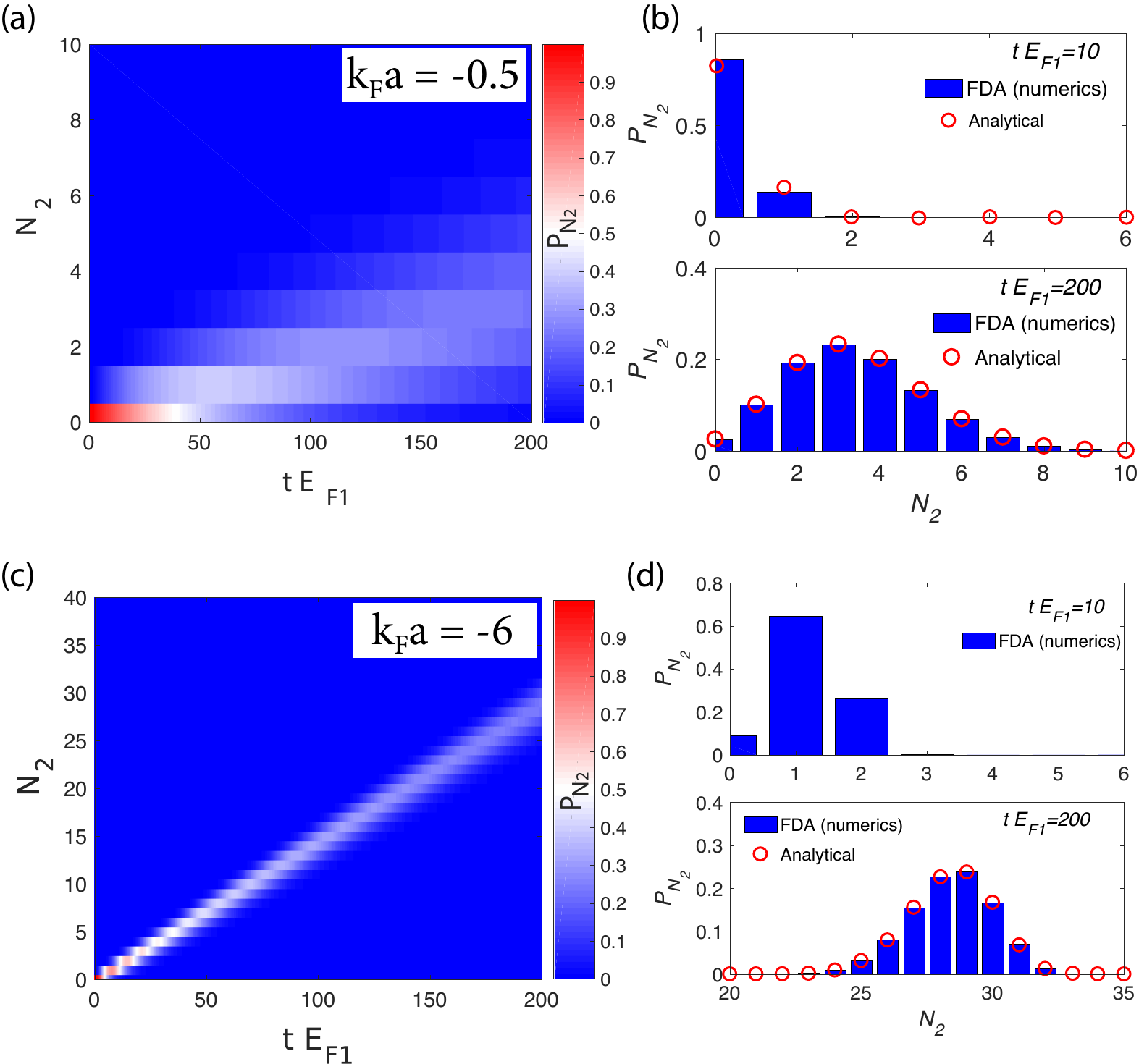}
\caption{\textbf{Full counting statistics.} Probability $P_{N_2}$ to measure $N_2$ atoms  for rotation angle $\theta=\pi/2$ and $E_{F2}/E_{F1}=0$ at zero temperature for $k_{F1}a=-0.5$  [Fig.~(a,b)], and $k_{F1}a=-6$ [Fig.~(c,d)].  The left panels show $P_{N_2}(t)$ as a function of the time $t E_{F1}$ and transferred spin number $N_2$ as obtained from the numerically exact FDA calculation.  The right panels show $P_{N_2}(t)$ at fixed times $t E_{F1}=10$ and $t E_{F1}=100$.  The numerical results are shown as blue bars and the  analytical prediction from Eq.~\eqref{LevitovChi} is shown as red circles.  
In the upper panel of Fig.~(d) we do not show analytical data as at such short times and strong interactions Eq.~\eqref{LevitovChi} becomes invalid.}\label{fig_4}
\end{figure}

In Fig.~\ref{fig_4}(a,c) we show the time evolution of $P_{N_2}$ for intermediate and strong interactions at zero temperature and  polarization angle $\theta=\pi/2$ as function of time $t$ and $N_2$. After sufficiently long times, the distribution is peaked around its mean value, and tracking the linear evolution of the mean with time makes the development of the steady spin current $J$ evident. However, what is the distribution of measured $N_2$  away from its mean? This question is studied in Fig.~\ref{fig_4}(b,d) where the distribution $P_{N_2}(t)$ obtained from the FDA is shown at fixed times $t E_{F1}=10$ and $100$ as blue bars.

The observed distributions  can again be studied in a wave packet picture. Over time wave packets reach the impurity and either remain in the original spin reservoir (only picking up a scattering phase shift) or they undergo a spin-changing collision.  For $N$ incoming particles within a time span $t$ there are $N$ trials to flip the spin. This line of argument leads us to Levitov's formula that describes fermions transmitted through a multi-channel barrier at zero temperature~\cite{Levitov1996,Nazarov2009},
\be\label{LevitovChi}
\ln \chi (\lambda,t)=t\int_0^{E_{F1}} \frac{dE}{2\pi\hbar} \ln[1+\Gamma(E)(e^{i\lambda}-1)].
 \ee
The data obtained from this expression, which is valid in the long-time limit, is shown as red circles in Fig.~\ref{fig_4}(b,d). The excellent agreement with the exact numerical result underlines the accuracy of the intuitive picture of wave packets of fermions scattering of the impurity and thereby flipping their spin with a finite probability. One can understand the FCS derived from Eq.~\eqref{LevitovChi} in various regimes analytically. For very weak coupling $|k_Fa| \ll 1$, where $\delta_k =-k a$, Eq.~\eqref{LevitovChi} reduces to the characteristic function of a Poisson distribution. At unitarity (where $a$ tends to infinity), $\delta_k =\pi/2$, $\Gamma(E)$ is independent of energy, and Eq.~\eqref{LevitovChi} becomes the characteristic function of a binomial distribution. Finally, in the regime in between, Eq.~\eqref{LevitovChi} represents  a superposition of binomial distributions; see App.~\ref{sec:appendixB}. We note that a finite number of impurities  leads to deviations from the FCS studied in this section, as discussed in  App.~\ref{app_CLT}.

\section{Non-equilibrium Orthogonality Catastrophe}\label{Sec:V}

So far we discussed how to employ the fermionic medium as a probe to study transport. However, our system also allows us to use the impurity as a probe of the many-body dynamics to study the `non-equilibrium orthogonality catastrophe' (NOC). In the `conventional' orthogonality catastrophe, as introduced by Anderson \cite{Anderson1967} and then extended to dynamics by Nozieres \etal \cite{nozieres_singularities_1969}, one considers a single-component Fermi gas in its ground state into which a scattering potential is suddenly introduced. This results in a quantum quench dynamics exhibiting a characteristic power law decay of the impurity Greens function \cite{nozieres_recoil_94,combescot_1971,yuval_exact_1970,nozieres_singularities_1969}. Extending this scenario, where the Fermi sea is initially in an equilibrium state, the \textit{non-equilibrium} orthogonality catastrophe refers to the situation where the system is initially in a non-equilibrium state. This scenario is realized in our setup since the system, despite being in a pure state, is initially not in its energetic ground state  of the non-interacting Hamiltonian $\hat H_0$ due to the large spin imbalance between the two  reservoirs. 

Previously it has been shown that quite generally the sudden introduction of a scattering potential into a system exhibiting  Fermi baths with multiple Fermi edges (in our case two), leads to a dynamical response of the system that features modified power-laws accompanied by exponential dampening \cite{abanin_tunable_2004, abanin_fermi-edge_2005,Gutman2011,Protopopov2013}. Here we bring together the results of these previous works as well as the study of subleading excitation branches and bottom-of-the-band dynamics introduced in Refs.~\cite{Knap2012,Schmidt2017}, and show how the dynamics can be observed in ultracold atom experiments. Combining both analysis allows us to analytically uncover a non-trivial connection between the decay of the Ramsey contrast and the tail of the FCS of spin transport. However, before we turn to the analytical analysis of the NOC, we consider the exact numerical solution of the problem and outline how it can be probed in experiments.

\subsection{Ramsey Spectroscopy}

One of the key signatures of the NOC is contained in the impurity Green's function that can be probed directly in Ramsey spectroscopy. To this end, the Fermi gas (in this section we allow for a finite Fermi energy $E_{F2}$) is first prepared in its initial non-equilibrium state by a spin rotation of the polarization angle $\theta$ leading to  $\ket{\psi_F}=\ket{\text{FS}_1}\otimes\ket{\text{FS}_2}$ (cf. Fig.~\ref{fig_1}). This is followed by a $\pi/2$ rf pulse acting on the impurity hyperfine states so that the initial state of the system reads $\ket{\Psi(0)}=\frac{1}{\sqrt{2}} (|u\rangle+|d \rangle)\otimes |\psi_F\rangle
$. After a time evolution for a time $t$, the expectation value of $\hat \sigma_x$ of the impurity spin is measured which yields the Ramsey signal \cite{Goold2011,Knap2012,Schmidt2017}
\be
\langle\hat{\sigma}_x\rangle=\textmd{Re}\langle\psi_F|e^{i \hat{H}_0 t } e^{-i \hat{H}_1 t}|\psi_F\rangle = \text{Re}S(t).
\ee
Moreover, by choosing the phase of the closing $\pi/2$ pulse, the  complex signal $S(t)$ can be measured \cite{Knap2012} which provides access to the full time-dependent response of the impurity spin \cite{Goold2011,Knap2012}.

As described in App.~\ref{sec:appendixA}, the overlap  $S(t)$ can be obtained numerically exactly using the FDA. The FDA allows us to map the calculation of many-body wave function overlaps onto the  evaluation of determinants in single-particle Hilbert space. For $S(t)$ one obtains
 \begin{equation}\label{FDAS}
S(t)=\bra{\psi_F}e^{i \hat{H}_0 t } e^{-i \hat{H}_1 t }\ket{\psi_F}=\det[\mathbbm{1}+\hat{n}(\hat{R}-\mathbbm{1})].
\end{equation}
Additional to $\mathbbm{1}=\textmd{diag}(1, 1)$, Eq.~\eqref{FDAS} contains two non-commuting block matrices: the two-component distribution matrix $\hat{n}=\textmd{diag}(\hat{n}_1, \hat{n}_2)$ that is diagonal in the (1,2)-basis ($\hat{n}_i=1/(e^{\beta(\hat h_{0,i}-\mu_i)}+1)$) and  the matrix $\hat{R}=\textmd{diag}(e^{i \hat{h}_{0,\uparrow} t } e^{-i\hat{ h}_{1,\uparrow} t }, \hat{1})$ which acts diagonally in the $(\uparrow,\downarrow )$-basis. Here $\hat{h}_{0,\uparrow},$ $\hat{ h}_{1,\uparrow}, $ and $\hat h_{0,i}$ are the single-particle representations of the many-body Hamiltonian $\hat{H}_{0,\uparrow}=\sum_{\veck }\epsilon_\veck \ced_{\veck\uparrow}\ce_{\veck\uparrow},$ $\hat{ H}_{1,\uparrow}=\hat{H}_{0,\uparrow} +    \frac{V_0}{\mathcal V}\sum_{\veck\vecq} \ced_{\veck\uparrow} \ce_{\vecq\uparrow}$ and $\hat{H}_{0,i=1/2}=\sum_{\veck} \epsilon_{\veck}\ded_{\veck i} \de_{\veck i},$ respectively.

\begin{figure}
	\centering
	\includegraphics[width=8.5cm]{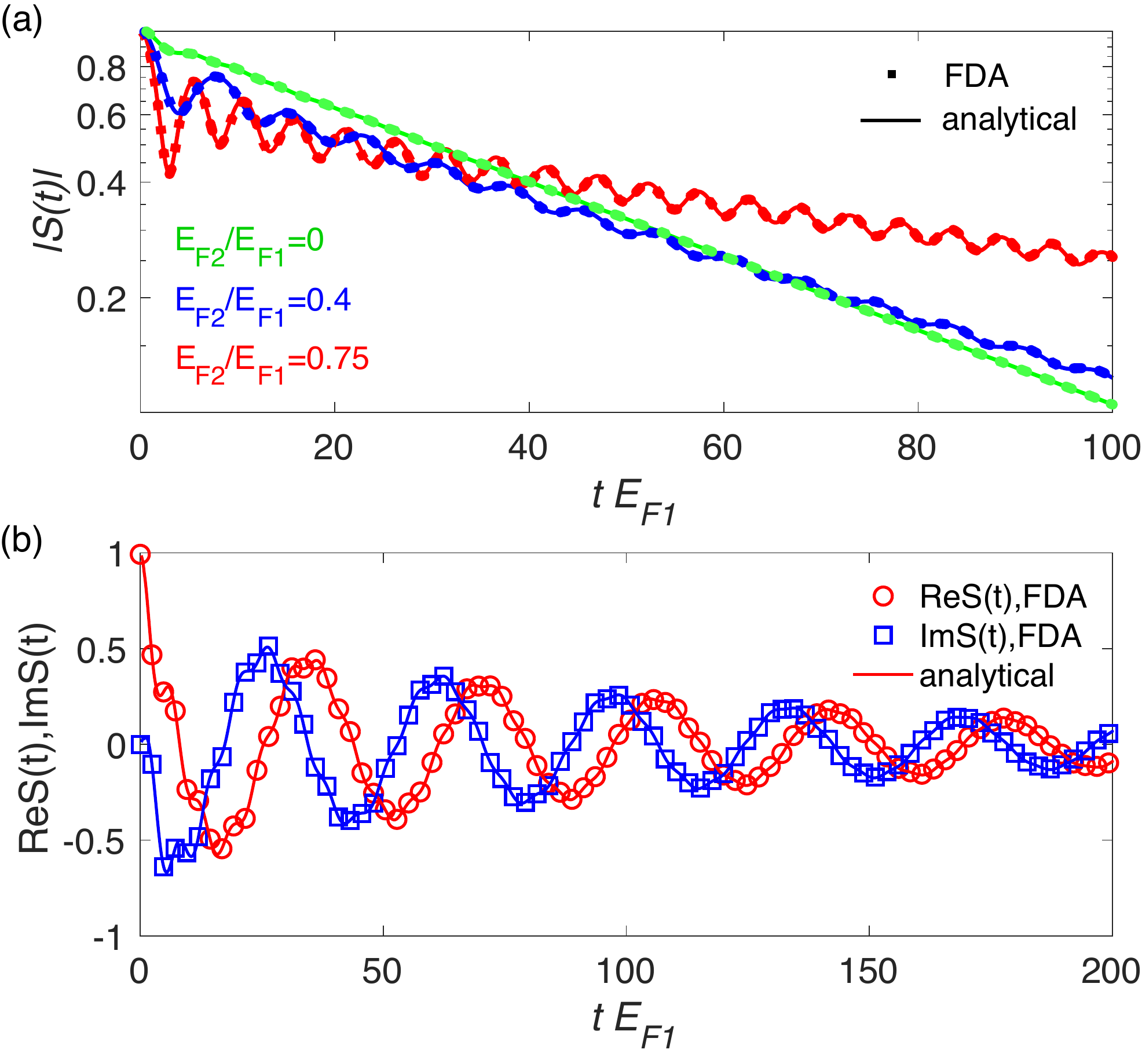}
	\caption{\textbf{Ramsey signal of the impurity.} (a) Ramsey contrast $|S(t)|$ for $\theta=3\pi/4$, scattering length $k_{F1}a=1.5$ and temperature $T=0$. Red, blue, green symbols correspond to the numerical FDA result for $E_{F2}/E_{F1}=0.75,$ $E_{F2}/E_{F1}=0.4$, and $E_{F2}/E_{F1}=0$, respectively, while the solid lines show the analytical prediction obtained from Eq.~\eqref{eq:BCSSot}, with  coefficients $C$ obtained from fits to the data. (b) Real and imaginary part of the Ramsey signal computed numerically using the FDA (symbols) for $\theta=3\pi/4,$ $k_{F1} a= 1.5$ and $E_{F2}/E_{F1} =0.75$. The solid lines are obtained from the asymptotic form, Eq.~\eqref{eq:BECSot}, using the coefficients $C$ as fit parameters. 
	}\label{fig_6}
\end{figure}

The time evolution of $S(t)$ at zero temperature  is shown in Fig.~\ref{fig_6} for  $E_{F1}\neq E_{F2}$. We find that $S(t)$ develops oscillations and an exponential damping at long times that persists even at zero temperature. In the conventional  orthogonality catastrophe, an exponential decay of $S(t)$ is  observed only for finite temperature $T>0$. There it indicates thermal decoherence due to the thermal occupation of single particle states given by the  Fermi distribution $n_\veck$. Thus one might be tempted to assume that the exponential decay observed in the NOC might be related to the development of a quasi-thermal state of the Fermi bath, which in turn induces quasi-thermal decoherence. However, as we have seen in the previous discussion that $n_{\sigma}(t,k)$ does not reach  a thermal state; see, e.g., Fig.~\ref{fig_2}. Therefore, the exponential decay of $S(t)$  must have a different origin and we will discuss below by analytical means.

\subsection{Analytical approach to the asymptotic behavior of $S(t)$ at zero temperature}\label{longtime}

\begin{figure}[t]
\centering
\includegraphics[width=7.5cm]{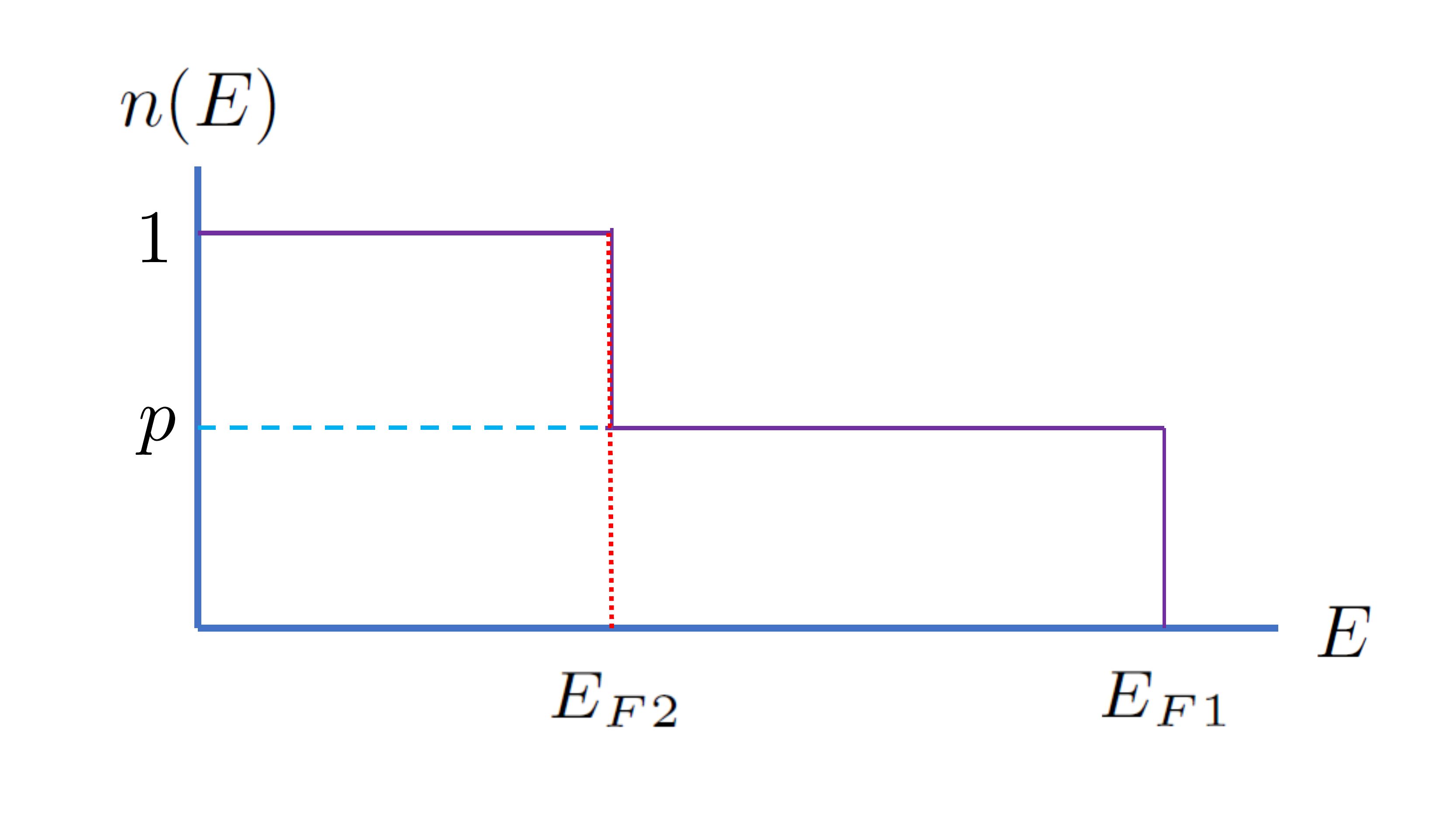}
\caption{\textbf{Effective two-step distribution function.} The expectation value of the effective single-particle occupation operator $\hat{n}(E)$ is given by a two-step function
$n(E)=(1-p) n_F(E-E_{F2}) +p n_F(E-E_{F1})$ with polarization $ p=\cos^2(\theta/2)$.
}\label{fig_momdist}
\end{figure}

Building on the insight from the numerically exact solution using the FDA, one may use the theory of Toeplitz determinants to derive analytical expressions that describe the exact dynamics with high accuracy also at intermediate times. To find such a description we first map the problem of an impurity interacting with two Fermi seas to the case of an impurity interacting with a single-component Fermi sea. To this end we express both $\hat{n}$ and $\hat{R}$  in the $(\uparrow,\downarrow)$ basis  using  a unitary transformation $(\ket{1},\ket{2})^T=\hat{U}(\ket{\uparrow}, \ket{\downarrow })^T$. A straightforward calculation (see App.~\ref{sec:appendixC}) shows that Eq.~\eqref{FDAS} can be expressed as 
\begin{equation}\label{SMapped}
S(t)=\det[1+(e^{i \hat{h}_{0,\uparrow} t } e^{-i\hat{ h}_{1,\uparrow} t }-{1}) \hat{n}(E)]
\end{equation}
where the associated single-particle occupation operator $\hat{n}(E)$ corresponds to the momentum distribution
 \be
n(E)=(1-p) n_F(E-E_{F2}) +p n_F(E-E_{F1}).\label{Eq:distribution}
\ee
This distribution is shown in Fig.~\ref{fig_momdist}. It exhibits  two Fermi surfaces at energies $E_{F1}$ and $E_{F2}$ and the polarization $p=\cos^2(\theta/2)$ determines the occupation of the middle plateau in $n(E)$. Using this transformation we have thus mapped the dynamics of the two-component Fermi gas onto the dynamics of a one component Fermi gas featuring two Fermi edges for which long-time solutions have been discussed in the literature \cite{Gutman2011,Protopopov2013}. 

In fact,  Eq.~\eqref{SMapped}  already allows one to qualitatively understand the source of the observed exponential decoherence persistent in the NOC at $T=0$. Indeed comparing Eq.~\eqref{SMapped} to the functional determinant formula Eq.~\eqref{KlichFormula} in App.~\ref{sec:appendixA} reveals that the dynamics is effective governed by a many-body density matrix that describes a \textit{single}-component Fermi gas not in a pure but in a \textit{mixed} state. It is the  classical nature of this state that provides the resource of exponential decoherence of the observed dynamics. We now turn  to support this  argument by a quantitative derivation.

Following Refs.~\cite{Knap2012,Schmidt2017}, we decompose $S(t)$ in terms of branches of different excitations of the Fermi system. These so-called `excitation branches' are
\begin{enumerate}[(i)]
\item
Particle-hole excitations near the two Fermi surfaces~[denoted as (FS1) and (FS2)].
\item Excitations from the bottom-of-the-band~(FB).
\item For $a > 0$, excitations involving the bound state~(BS).
\end{enumerate}

We  focus first on the `attractive  interaction regime', where the scattering length $a$, as determined by the low-energy expansion of the phase shift $\delta_k=-k a$, is negative, $a<0$. Using the formulation in terms of a single Fermi sea, the asymptotic behavior of $S(t)$  can be organized as
\begin{align}
S(t)=\sum_{n_1+n_2+n_3=0}  & C_{n1,n2,n3} \,\, e^{- i  \kappa_0 t}\nonumber\\
&\cdot \,\,S_{n_1}^{(FS1)}(t) S_{n_2}^{(FS2)}(t) S^{(FB)}_{n_3}(t).
\label{eq:BCSSot}
\end{align}
Here the subscript $n_{i=1,2}$ refers to the number of particles added to or removed from the first and second Fermi
edge, respectively, while $n_3<0$ refers to the number of particles removed from the bottom of the Fermi sea.  Particle number conservation imposes the constraint $n_1+n_2+n_3=0$.

While the coefficients $C_{n1,n2,n3} $ depend on the microscopic details, the other contributions in Eq.~\eqref{eq:BCSSot} can be cast in analytical form. The complex-valued constant $\kappa_0$ is, for instance, given by
\cite{Gutman2011,Protopopov2013} (see also App.~\ref{sec:appendixD})
\begin{align}
\kappa_0= \Delta E_0 - i \gamma=&i\int^{\infty}_0 \frac{d E}{2 \pi} \ln[ 1+{n}(E)( e^{2 i \delta(E)}-1)]\nonumber\\
=&- \int^{E_{F2}}_0 \frac{dE}{\pi} {\delta}(E)-\int^{E_{F1}}_{E_{F2}} \frac{dE}{\pi} {\delta}_{\textmd{eff}}(E).
\label{eq:DeltaE}
\end{align}
Here the first term of the last expression is obtained from the integration from $0\ldots E_{F2}$ where $n(E)=1$. The second term originates from the remaining integration region  $E_{F2}\ldots E_{F1}$ where $n(E)=p<1$. It involves the effective phase shift defined by
\be
{\delta}_{\textmd{eff}}(E)&=&-\frac{i}{2} \ln\big[1+p (e^{2 i {\delta}(E)}-1) \big],
\ee
and represents a  generalization of Fumi's theorem (that expresses the ground state energy as a sum over scattering phase shifts~\cite{Mahan2000,Giamarchi2004}) of the conventional OC to the  case of spin-flip interactions.

The analytical calculation of the time-dependent factors $S_i(t)$ in Eq.~\eqref{eq:BCSSot} is challenging in a naive bosonization approach. Instead, the use of Szeg\H{o} formula~\cite{Basor1994,Fisher,Hassler2009,Deift2011} to second order allows one to approach the problem. Indeed, Gutman and coworkers showed that the contributions involving exclusively particle-hole fluctuations  close to the two Fermi edges  are given by~\cite{Gutman2011,Protopopov2013}
\begin{eqnarray}\label{phcon}
S_0^{(FS1)}(t)&\propto& t^{-\left(\frac{\tilde{ \delta}_1 }{\pi}\right)^2}\\
S_0^{(FS2)}(t)&\propto& t^{-\left(\frac{\tilde{ \delta}_2}{\pi}\right)^2}.
\end{eqnarray}
These expressions represent the Fermi edge singularities and exhibit a non-trivial power-law behavior with exponents determined by (see App.~\ref{sec:appendixD})
\be\label{exponents}
\tilde{\delta}_1&=&{\delta}_{\textmd{eff}}(E_{F1}-0^+)\\
\tilde{\delta}_2&=&{\delta}(E_{F2}+0^+)-{\delta}_{\textmd{eff}}(E_{F2}-0^+).
\ee

Generalizing this analysis to the case where $n$ particles are added or removed from the Fermi edges at $E_{F1}$ and $E_{F2}$ allows one to describe analytically not only the long- but also the intermediate-time dynamics with high accuracy \cite{Gutman2011,Protopopov2013}.  In App.~\ref{sec:appendixD} we provide a detailed derivation that leads to the expressions
\begin{align}\label{bbformula}
S_n^{(FS1)}(t)&\propto e^{- i n E_{F1} t}  \Big(\frac{1}{ t}\Big)^{\left(\frac{\tilde{\delta}_1}{ \pi}-n\right)^2},\nonumber\\
S_n^{(FS2)}(t)&\propto e^{- i n E_{F2} t} \Big( \frac{1}{ t}\Big)^{\left(\frac{\tilde{\delta}_2}{ \pi}-n\right)^2}
\end{align}
that are valid in the zero-temperature limit. Note that here we include the phase factors that depend on the Fermi energies  into the definitions of $S_n^{(FS1,2)}(t)$, which is a different convention compared to  Ref.~\cite{Schmidt2017}. To reflect this choice we  introduced the subindex $n=0$ in $ \kappa_0$ given by Eq.~\eqref{eq:DeltaE}.

A further contribution which has so far not been studied in the context of NOC dynamics originates from processes where particles are excited from the bottom of the band to  the two edges of the Fermi sea, leaving holes behind. The corresponding contribution can be found from few-body theory and reads~\cite{Knap2012,Schmidt2017}
\be\label{FBEQ}
S^{(FB)}_{n}\propto\Big[\int^{\infty}_0 \frac{d E}{\sqrt {E} }\sin^2 \delta(E) e^{i E t}\Big]^{-n},
\ee
with $n\leq 0$. 

We now turn to  the  interaction regime for $a>0$, where a weakly-bound state of energy $E_b<0$ exists. Here, the overlap $S(t)$ can be expressed as
\begin{align}\label{eq:BECSot}
S(t)=&\sum_{n_1+n_2+n_3+n_4=0} C_{n_1,n_2,n_3,n_4} \,\,e^{- i\kappa_0 t}  \nonumber\\
&\quad\quad\quad\cdot S_{n_1}^{(FS1)}(t) S_{n_2}^{(FS2)}(t) S^{(FB)}_{n_3}(t)S^{(BS)}_{n_4}(t).
\end{align}
The index $n_4$ takes on values $0$ or $1$, depending on whether the bound state is occupied or empty; i.e., $S^{(BS)}_{1}=e^{- i E_b t}$ or $S^{(BS)}_{0}=1$, respectively. In Fig.~\ref{fig_6}(b) we compare the analytical expression  to the numerical results. Here the coefficients $C$ serve as fit parameters and we keep only  leading contributions with $\sum_i |n_i|\leq 2$. We find that the asymptotic form reproduces the exact numerical  results with remarkable precision down to  small evolution times. Here the superposition of  oscillating factors from  bottom-of-the-band contributions (given by Eq.~\eqref{FBEQ}), bound-state (proportional to $e^{i E_b t}$) and Fermi surface contributions (proportional to $\sim e^{-i n E_{Fi}t}$, c.f.~Eq.~\eqref{bbformula}) results in the oscillations visible in Fig.~\ref{fig_6}.

\subsection{Role of finite temperature} \label{TSection}

In the previous discussion we have found that a key signature of the non-equilibrium orthogonality catastrophe is the exponential decay of $|S(t)|\sim e^{-\gamma t}$ that is present even at zero temperature. We now focus on the temperature dependence of the  decay rate $\gamma$. Using the Szeg\H{o}  theorem for the asymptotic properties of Toeplitz determinants, we find  
\be\label{eq:asymptotic}
 \gamma=- \textmd{Re} \int^{\infty}_0 \frac{d E}{2 \pi} \ln[ 1+n(E)( e^{2 i {\delta}(E)}-1)].
\ee
 In this expression, which follows from Eq.~\eqref{eq:DeltaE} (for details see App.~\ref{sec:appendixD}), we take into account  the energy-dependent phase shift $\delta(E)$ and the temperature-dependent distribution function $n(E)$  given by Eq.~\eqref{Eq:distribution}.

\begin{figure}[t]
\centering
\includegraphics[width=8.5cm]{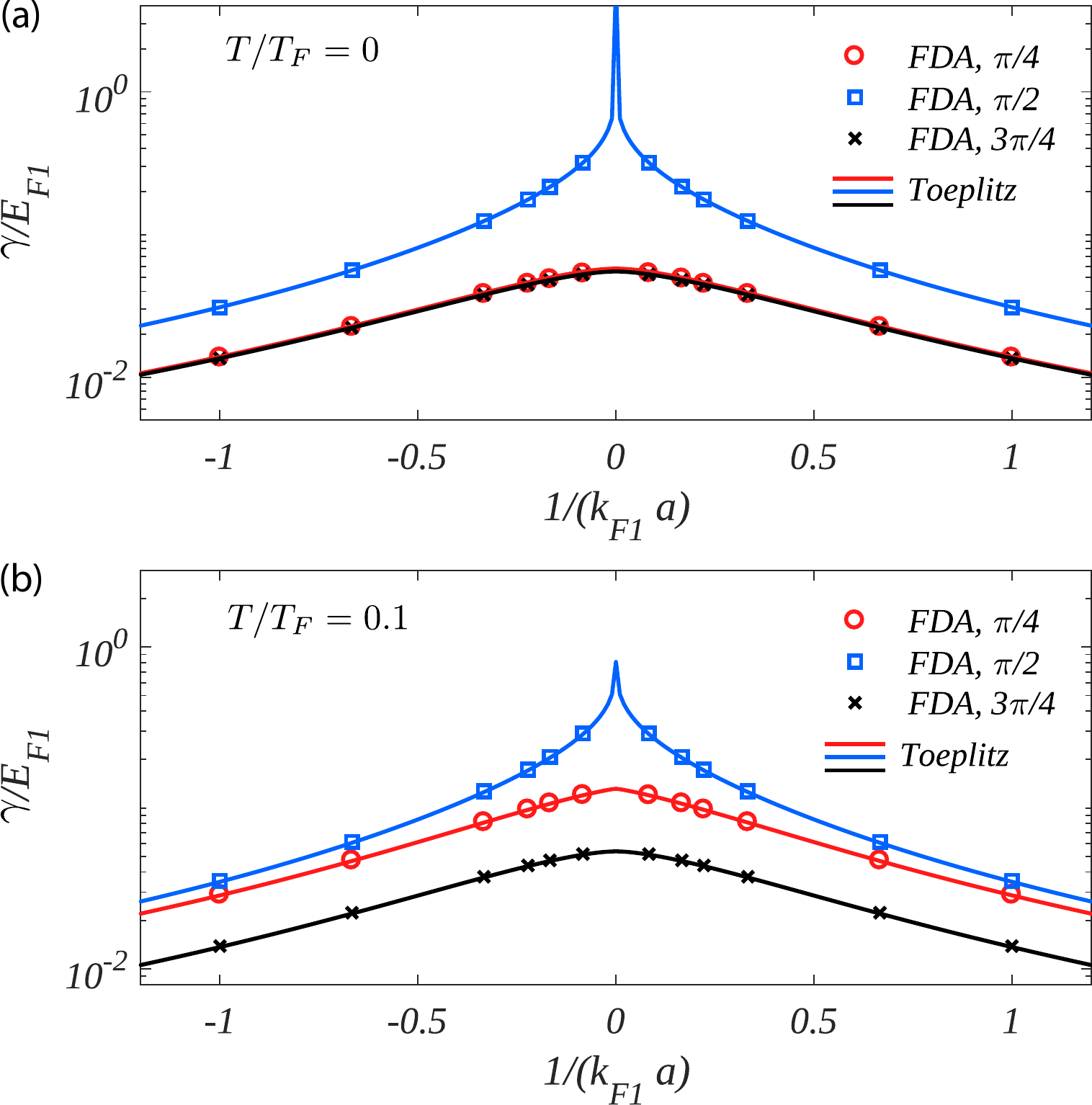}
\caption{ \textbf{Asymptotic decay rate $\gamma$  of the Ramsey signal.} The decay rate $\gamma$ is shown as a function of scattering length $k_{F1} a$ for three different values of the polarization angle $\theta=\{ \pi/4, \pi/2, 3\pi/4 \}$, an initially empty second reservoir $E_{F2}/E_{F1}=0$, and temperatures (a) $T/T_{F1}=0$ and (b) $T/T_{F1}=0.1$. The solid lines  are obtained from Eq. \eqref{eq:asymptotic}.
}\label{fig_7}
\end{figure}

In  Fig.~\ref{fig_7} this analytical result is compared  to the decay rate obtained from fitting $|S(t)|\sim e^{-\gamma t}$ to the exact FDA results at long times. We find excellent agreement between the numerical FDA data and the analytical expression both when studying  the $\theta$ and $1/k_{F1}a$ dependence of $\gamma$ for the two temperatures $T/T_F=0$ and $T/T_F=0.1$.

Using the relation  $\textmd{Re}\ln[ 1+p( e^{2 i \delta(\varepsilon)}-1)]=\textmd{Re}\ln[ 1+(1-p)( e^{2 i \delta(\varepsilon)}-1)]$  one finds from the analytical prediction Eq.~\eqref{eq:asymptotic} that the decay rate is symmetric with respect to $p=1/2$ at zero temperature, as shown by the comparison of $p=\cos^2(\pi/4)$ and $p=\cos^2(3\pi/4)$ in Fig.~\ref{fig_7}(a).  At finite temperatures this symmetry is absent and, as shown in Fig.~\ref{fig_7}(b), we find that $p>1/2$ exhibits a larger decay rate than $p<1/2$. The reason for the different decay rates lies in the fact that it is spin-\textit{conserving} collisions within a reservoir, as determined by the diagonal element of the scattering matrix in Eq.~\eqref{SpinPumpHamiltonian},  that give rise to additional thermal decoherence; and since we have chosen the reservoir $1$ to have a larger occupation, polarizations $p=\cos^2(\theta/2)<1/2$ will give a larger decoherence rate compared to $p>1/2$.

\subsection{Relation between Ramsey interferometry and the FCS of spin flips}

It turns out that the decay rate of the Ramsey signal has a remarkable relation to the FCS of spin flips. Specifically, we find that the decay rate $\gamma$ in Eq.~\eqref{eq:asymptotic} and the FCS at zero temperature and $E_{F2}=0$, as described by the time-dependent generating function $\chi$ in Eq.~\eqref{LevitovChi}, are related by (see App.~\ref{app.ramsey.FCS}):
\begin{equation}
|S(t)|\to e^{\frac{1}{2} \ln \chi(e^{i\lambda}\to0 )} 
\end{equation}
From this equation directly follows the relation Eq.~\eqref{eq:relationSP}, $|S(t)|\sim \sqrt{P_{N=0}(t)}$, which holds up to logarithmic corrections.

This relation implies that the Ramsey decoherence is given by the square root of the probability  of having no spin flips in the time interval $0\ldots t$, which fits the notion  of $P_{N=0}(t)$  as an `idle-time probability', similar to the emptiness probability discussed in other contexts \cite{Abanov2003,Franchini2005}. Thus the Ramsey signal is related directly to the FCS and thus the intrinsic quantum noise in the number  of observed spin flips. Therefore, the decay of the Ramsey signal can serve as an indirect probe of the  tail of the FCS at low particle number. 

The relation Eq.~\eqref{eq:relationSP} can be understood as follows: the Ramsey contrast $|S(t)|$ is determined by the overlap of many-body states. When the spin of one of the fermions is flipped, a state of the Fermi system results that is orthogonal to the initial state, leading to a vanishing Ramsey contrast. Therefore, finding a finite Ramsey contrast requires configurations that have no fermion spin flipped. The probability of such a configuration is $P_{N=0}$. The Ramsey contrast $|S(t)|$ measures, however, an \textit{amplitude} (cf. Eq.~\eqref{FDAS}) so that $|S(t)|$ is proportional to $\sqrt{P_{N=0}}$.

\section{Conclusion and discussion}\label{Sec:VI}

In this work we proposed an ultracold atom experiment where  impurities are coupled to a spin-imbalanced two-component Fermi gas. The setup  allows one to study fundamental relations between quantum fluctuations in transport and dephasing dynamics. Specifically, we showed that applying rf pulses to the Fermi system provides a means  to realize initial non-equilibrium spin populations that are required to study spin transport. Based on a functional  determinant approach we explored the full counting statistics of the spin flips that accompany the spin current generated in our setup. 

Furthermore, we showed that the dynamics of the many-body wavefunction can be explored using Ramsey interferometry. This opens the path toward the study of the non-equilibrium orthogonality catastrophe (NOC) with ultracold quantum gases. The NOC  is characterized by a  decay of the Ramsey signal which is exponential although the system is initially in a pure quantum state, and one thus might have naively expected a power-law decay as obtained for the Fermi edge singularity. By mapping the problem onto a multi-Fermi edge scenario in energy space, we obtained analytic predictions for the long-time impurity response and, in particular, for its exponential decay rate. This allowed us to uncover a  relation between the FCS of spin flips and the rate at which the Ramsey contrast of the impurity decays. In this work we considered local quench-type dynamics, in which the impurity strength is changed only once. In order to explore a  broader class of non-equilibrium phenomena, one may  include multiple quenches of the scattering phase shift. Mathematically handling such multiple discontinuities will require a further generalization of the theory of Toeplitz determinants with Fisher-Hartwig singularities~\cite{Protopopov2013}. In this respect ultracold atom experiments might provide a quantum tool to explore mathematical problems for which solutions have yet to be found. 

Moreover, in the present work we did not attempt to explore ways to explicitly control the FCS of spin flips. In this regard it will be interesting to study whether it is possible to suppress fluctuations imprinted in the FCS by controlling and manipulating the impurity potential similarly to the realization of a source of pure single-particle spin transmission~\cite{Ivanov1997,Levitov1996,Keeling2006}. Finally, it has recently been shown  that von Neumann and the Renyi entanglement entropies can be expressed in terms of even order cumulants \cite{Klich2009,Song2011}. The fact that the full counting statistics contains the information about moments of arbitrary order  thus suggests that our proposed  scheme might enable one to further explore the relation between entanglement dynamics and  full counting statistics in cold atom experiments.

\textbf{\textit{Acknowledgments}}
We thank Rudolf Grimm and Dimitri Abanin for fruitful discussions. We acknowledge support from Harvard-MIT CUA, NSF Grant No. DMR-1308435, AFOSR-MURI: Quantum Phases of Matter, AFOSR-MURI: Photonic Quantum Matter, award FA95501610323. J.-S. Y. was supported by the Ministry of Science and Technology, Taiwan~(Grant No. MOST 104-2917-I-564-054). M. K. acknowledges support from the Technical University of Munich - Institute for Advanced Study, funded by the German Excellence Initiative and the European Union FP7 under grant agreement 291763.

\onecolumngrid
\newpage
\appendix
\section{Functional determinant approach and solution of the single-particle problem}\label{sec:appendixA}

For any bilinear many-body operator $\hat{X}_\alpha= \sum_{ij} \bra{i}\hat x_\alpha \ket{j}\hat{c}^{\dagger}_i  \hat{c}_j$, we can make use of  the identity 
\begin{align}\label{KlichFormula}
\langle e^{\hat{X}_1} \cdots e^{\hat{X}_N}\rangle=\text{Tr}[\hat\rho e^{\hat{X}_1} \cdots e^{\hat{X}_N}] =\textmd{det}(1-\hat{n}+\hat{n }e^{\hat{x}_1 }\cdots e^{\hat{x}_N}),
\end{align}
with $\hat \rho$ the density matrix and $\hat{n}$ denotes the corresponding single-particle occupation operator.  Hence, $S(t)$ can be expressed as
 \be
S(t)= \langle e^{i \hat H_0 t} e^{-i \hat H_1 t}\rangle =\det[\hat{1}- \hat{n}+\hat{n} e^{i \hat{h}_0 t } e^{-i\hat{ h}_1t }],
\ee
where $\hat{h}_0$ and $\hat{ h}_1 $ are the single-particle Hamiltonians in the absence and presence of impurity, respectively.
 To evaluate the functional determinant numerically, we work in the basis of single-particle eigenstates of $\hat{h}_0$ and $\hat{ h}_1 $.

 To this end, we solve the single particle problem in the presence of localized impurity. The  Schr\"{o}dinger equation for the two-component host fermions is given by
\be
 \begin{pmatrix}  -\frac{\nabla^2}{2m}+V(\vecr)\cos^2(\frac{\theta}{2}) & V(\vecr)\cos(\frac{\theta}{2})\sin(\frac{\theta}{2}) \\
V(\vecr)\cos(\frac{\theta}{2})\sin(\frac{\theta}{2})   &  -\frac{ \nabla^2}{2m}+V(\vecr)\sin^2(\frac{\theta}{2})
 \end{pmatrix}\begin{pmatrix}  \psi_1(\vecr)  \\
   \psi_2(\vecr)
 \end{pmatrix}=E\begin{pmatrix}  \psi_1(\vecr)  \\
   \psi_2(\vecr)
 \end{pmatrix}\label{eq:Schrodinger}
\ee
where $V(\vecr)$ is the short-range  potential. For our numerics we consider a finite system confined in a sphere of radius $R$ chosen large enough so that finite size corrections are negligible. For short-range interactions only the s-wave components of the scattering wave functions experience a phase shift. Defining the radial wavefunction $\phi_n(r)$  via $\psi_n(\vecr)=\phi_n(r)/(\sqrt{4\pi}r)$ with nodal quantum number $n$, Eq. \eqref{eq:Schrodinger} is expressed as a  radial one-dimensional Schr\"{o}dinger equation. The interaction between the impurity and itinerant fermions is fully characterized by the scattering length $a$ with the s-wave scattering phase shift given by ${\delta}_k =- \tan^{-1} k a$.

When the host fermions do not interact with the impurity, the eigenfunctions are given by
\be
\phi_{1, n}(r)=\sqrt {\frac{2}{R}} \sin (k_n r ) \otimes |1\rangle ,\; \phi_{2, n}(r)=\sqrt {\frac{2}{R}} \sin (k_n r ) \otimes |2\rangle,
\ee
with the boundary condition $k_n R= n \pi.$

In presence of the scattering potential, Eq.~\eqref{eq:Schrodinger} has  solutions with energies $E_n={k'_n}^2/2m $ that are determined by $k'_n R+  \delta_{k'_n}= n \pi$ and eigenstates 
\be
\phi_{ n}(r)=A_n \sqrt {\frac{2}{R}} \sin (k'_n r + \delta_{k_n}) \otimes \left[\cos\left(\frac{\theta}{2}\right)\ket{1} +\sin\left(\frac{\theta}{2}\right)\ket{2}\right]
\ee
where $A_n= 1/\sqrt {1+\frac{\sin 2 \delta_{k'_n}}{2 k'_n R}}$. There exists also a second set of solutions that is given by the noninteracting solutions determined by $E_0(n)=(k_n)^2/(2m) $  and
\be
\phi_{0, n}(r)=\sqrt {\frac{2}{R}} \sin (k_n r ) \otimes \left[\sin\left(\frac{\theta}{2}\right)\ket{1} -\cos\left(\frac{\theta}{2}\right)\ket{2}\right] .
\ee
Finally, for $a>0$  a bound state exists  with energy $E_{b}= -1/(2m a^2) $ and eigenfunction
\be
\phi_{b}(r) =A_b e^{-r/a} \otimes \left[\cos\left(\frac{\theta}{2}\right)|1\rangle +\sin\left(\frac{\theta}{2}\right)|2\rangle\right].
\ee
Here $A_b=\sqrt{\frac{2}{ a }}$ up to corrections that vanish as $R\to\infty$.

\section{Spin flip probability $\Gamma(E)$}\label{app:SpinFlipRate}
Here we derive an analytical expression for the spin flip probability $\Gamma(E)$ given by Eq.~\eqref{SpinFlipRate} in the main text. Scattering occurs only between fermions in their $|\uparrow\rangle$ spin state and the impurity in the $|u\rangle$ state:
\be\label{appB1}
|\uparrow\rangle\otimes|u\rangle&\mapsto& e^{i 2 \delta(E)}|\uparrow\rangle\otimes|u\rangle\nonumber\\
|\uparrow\rangle\otimes|d\rangle&\mapsto&|\uparrow\rangle\otimes|d\rangle\nonumber\\
|\downarrow\rangle\otimes|u\rangle&\mapsto& |\downarrow\rangle\otimes|u\rangle\nonumber\\
|\downarrow\rangle\otimes|d\rangle&\mapsto&|\downarrow\rangle\otimes|d\rangle,
\ee
where $\delta(E)$ is the energy-dependent phase shift.

Initially we apply a spin rotation  such that each fermion is prepared in a superposition state
\be
|1\rangle&=\cos(\theta/2)|\uparrow\rangle-\sin(\theta/2)|\downarrow\rangle,\\
|2\rangle&=\sin(\theta/2)|\uparrow\rangle+\cos(\theta/2)|\downarrow\rangle.
\ee
When the impurity is switched into the interacting state $|u\rangle$, the bath fermions, now prepared in states $|1\rangle$ to $|2\rangle$, undergo spin flip interactions. Using Eq.~\eqref{appB1} this scattering process is described by
\be
|1\rangle&\mapsto&\cos(\theta/2)e^{i 2 \delta(E)}|\uparrow\rangle-\sin(\theta/2)|\downarrow\rangle,\\
|2\rangle&\mapsto&\sin(\theta/2)e^{i 2 \delta(E)}|\uparrow\rangle+\cos(\theta/2)|\downarrow\rangle
\ee
When rewriting this process in the basis of $|1\rangle,|2\rangle$ 
\be
|1\rangle&\mapsto& [e^{i 2 \delta(E)}\cos^2(\theta/2)+\sin^2(\theta/2)]|1\rangle\nonumber\\&&+(e^{i 2 \delta(E)}-1)\sin(\theta/2)\cos(\theta/2)|2\rangle,\\
|2\rangle&\mapsto&(e^{i 2 \delta(E)}-1)\sin(\theta/2)\cos(\theta/2)|1\rangle\nonumber\\&&+[e^{i 2 \delta(E)}\sin^2(\theta/2)+\cos^2(\theta/2)]|2\rangle,
\ee
 one can directly read of the spin flip probability 
 \begin{equation}
 \Gamma(E)=|(e^{i 2 \delta(E)}-1)\sin(\theta/2)\cos(\theta/2) |^2=\sin^2 \theta \sin^2 \delta(E).
 \end{equation}

\section{Non-equilibrium momentum population and FCS in a given energy interval}\label{sec:appendixB}

Eq.~\eqref{LevitovChi} shows that the FCS of the total number of spin flips is determined as a sum involving the scattering probability for each momentum mode of the fermions. Hence, according to this expression, the FCS of spin flips in each individual momentum mode gives rise to a binomial distribution. In this appendix, we show that this argument is indeed confirmed by exact numerical simulation using the FDA.

\begin{figure}[b]
	\centering
	\includegraphics[width=4.5cm]{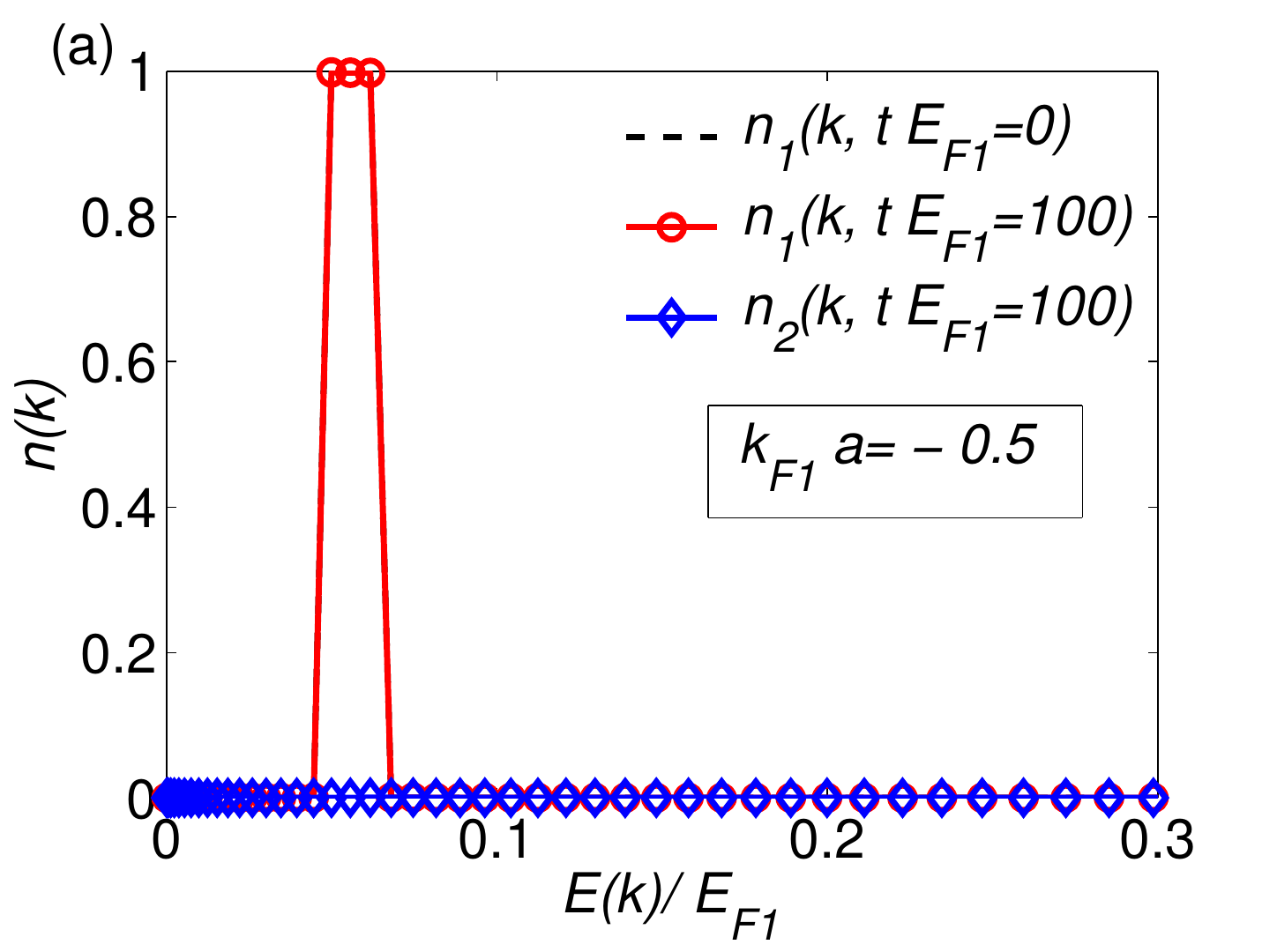}
	\includegraphics[width=4.5cm]{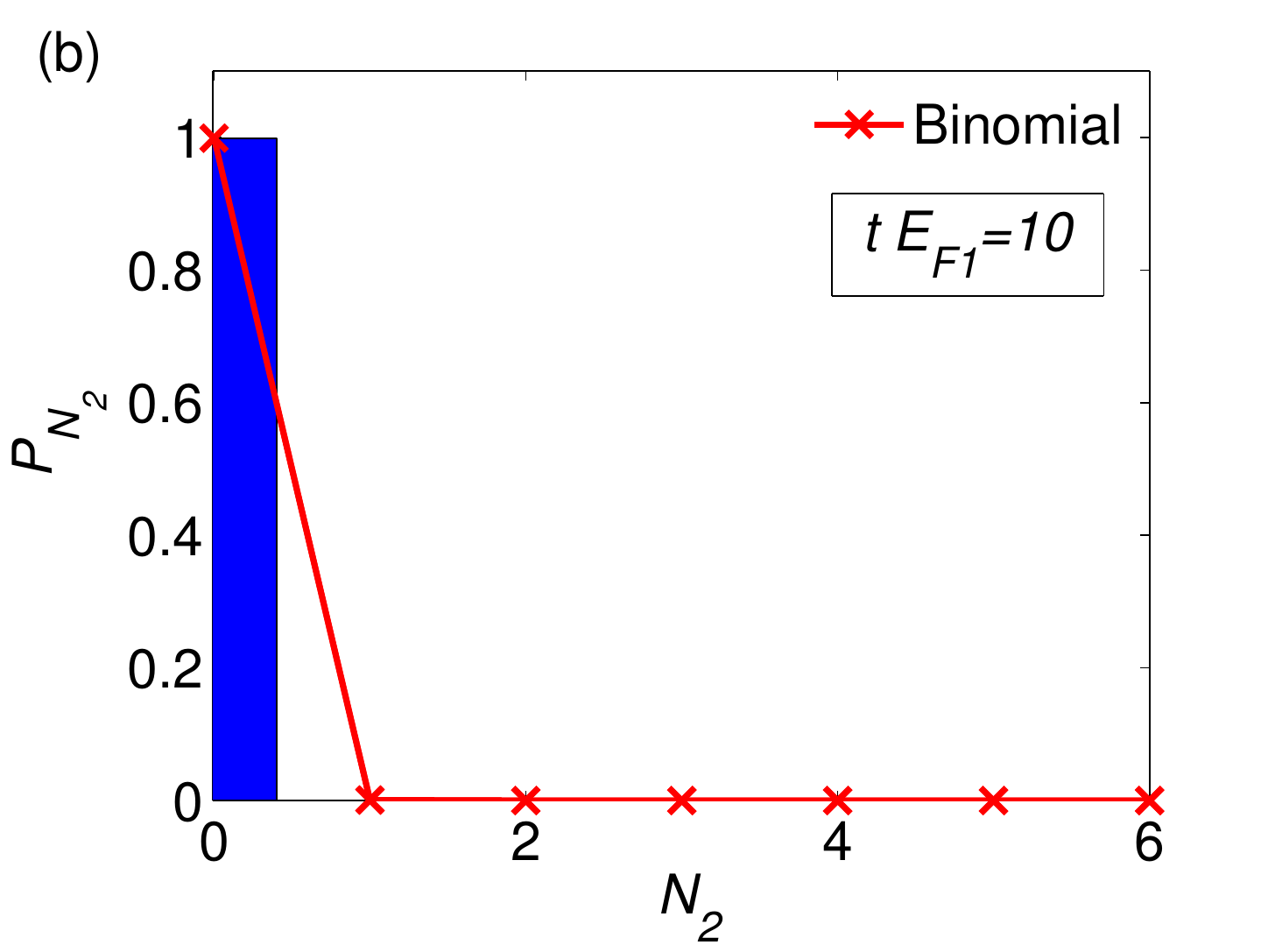}
	\includegraphics[width=4.5cm]{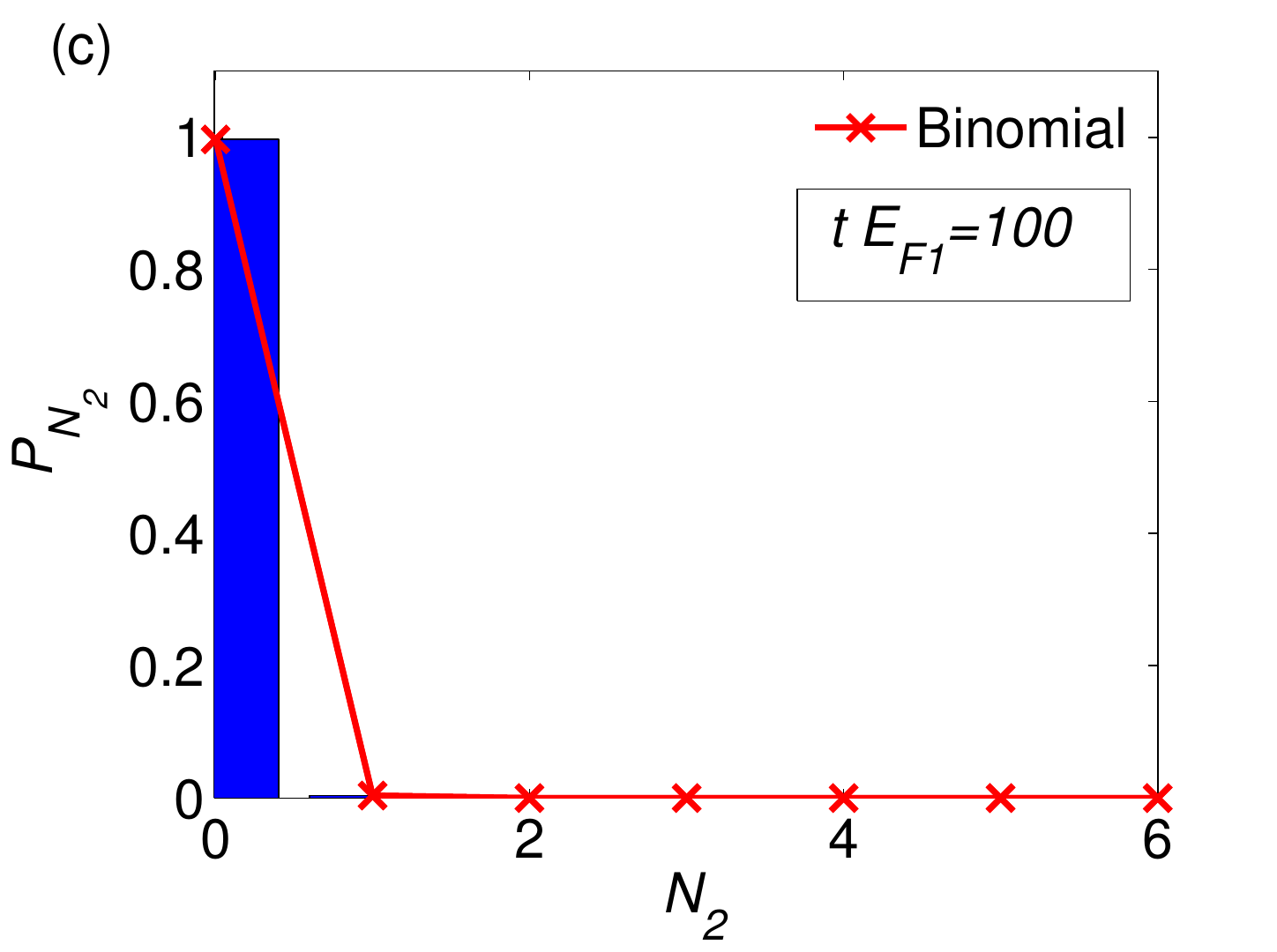}
	\includegraphics[width=4.5cm]{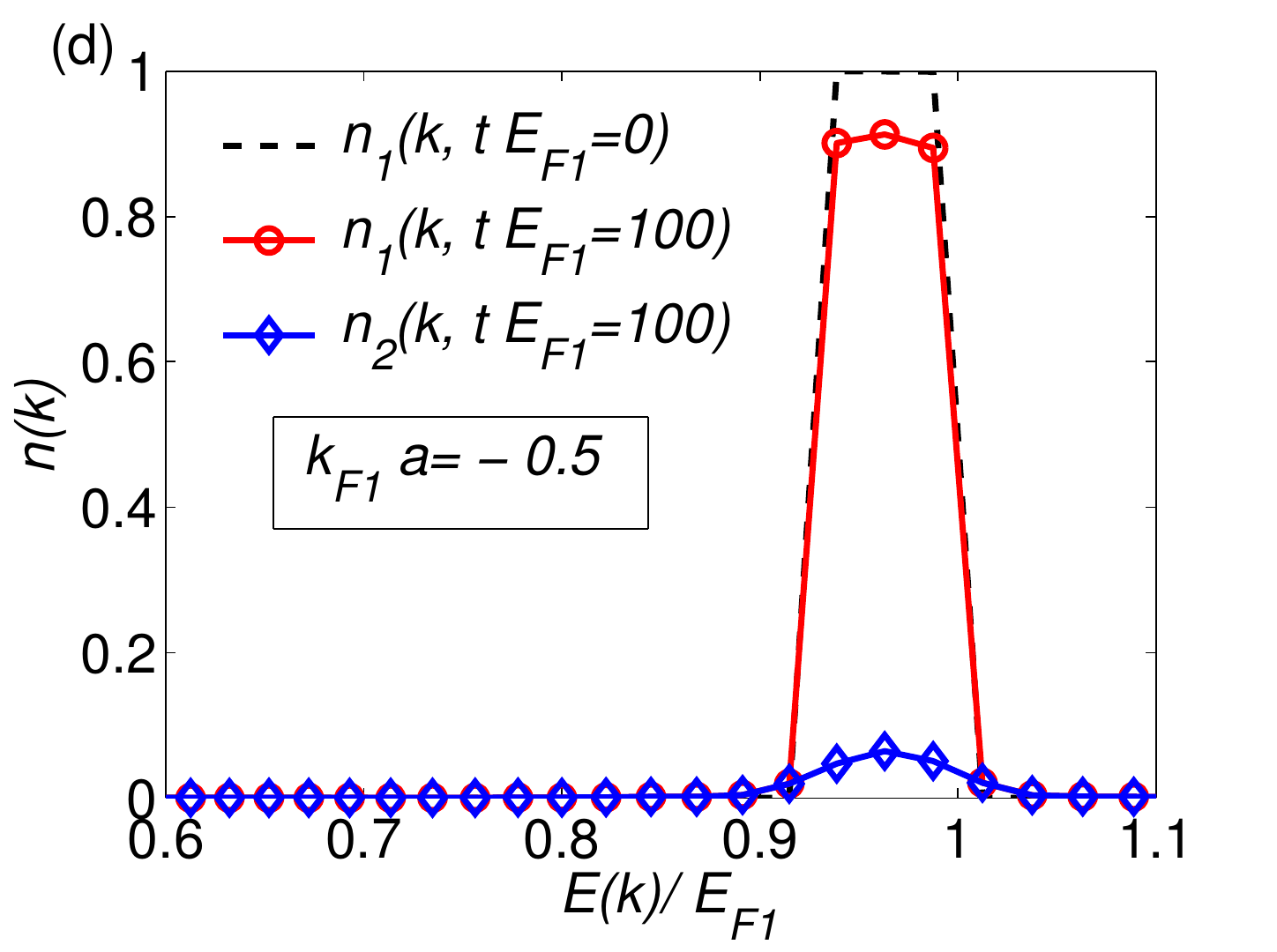}
	\includegraphics[width=4.5cm]{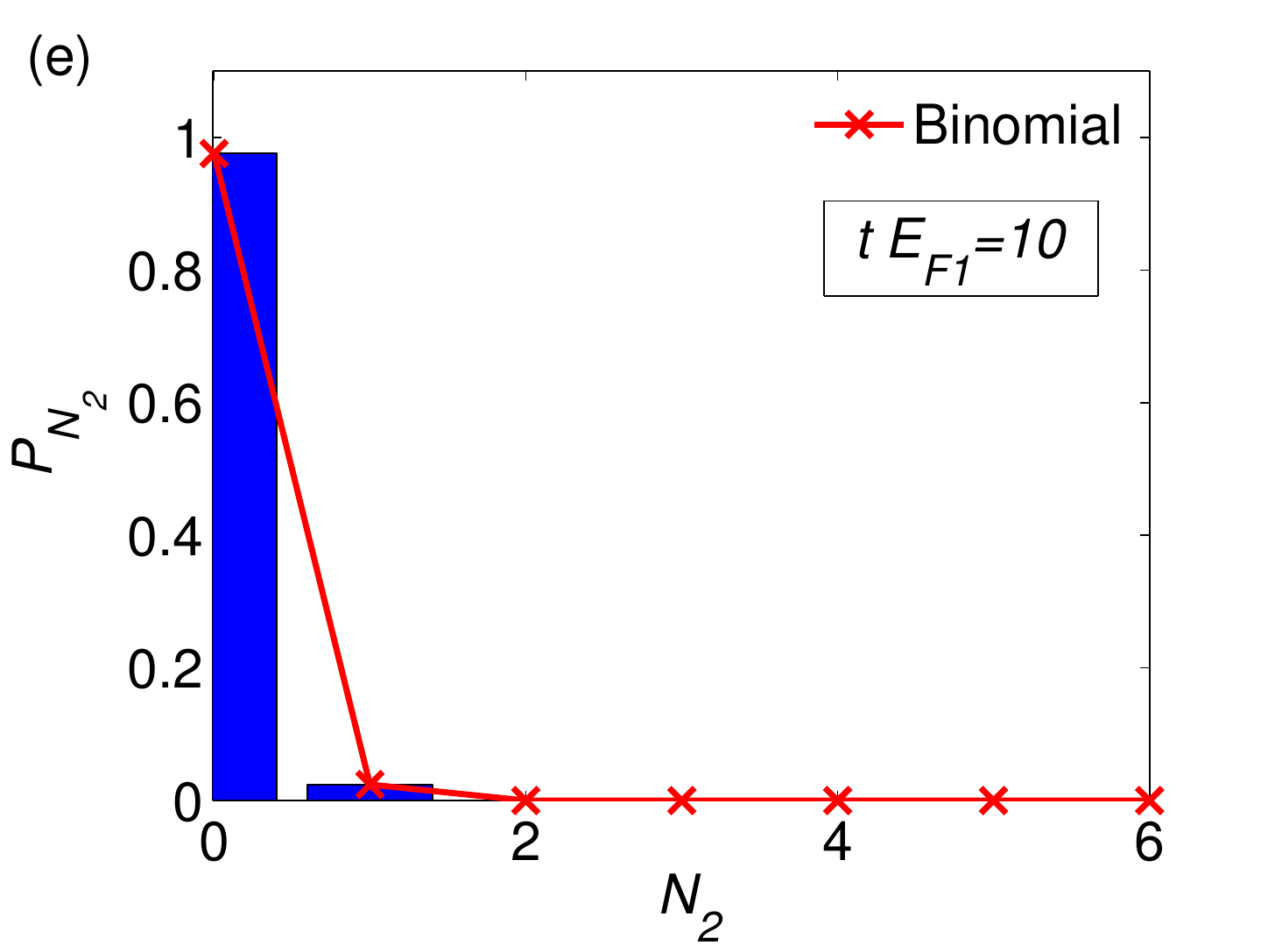}
	\includegraphics[width=4.5cm]{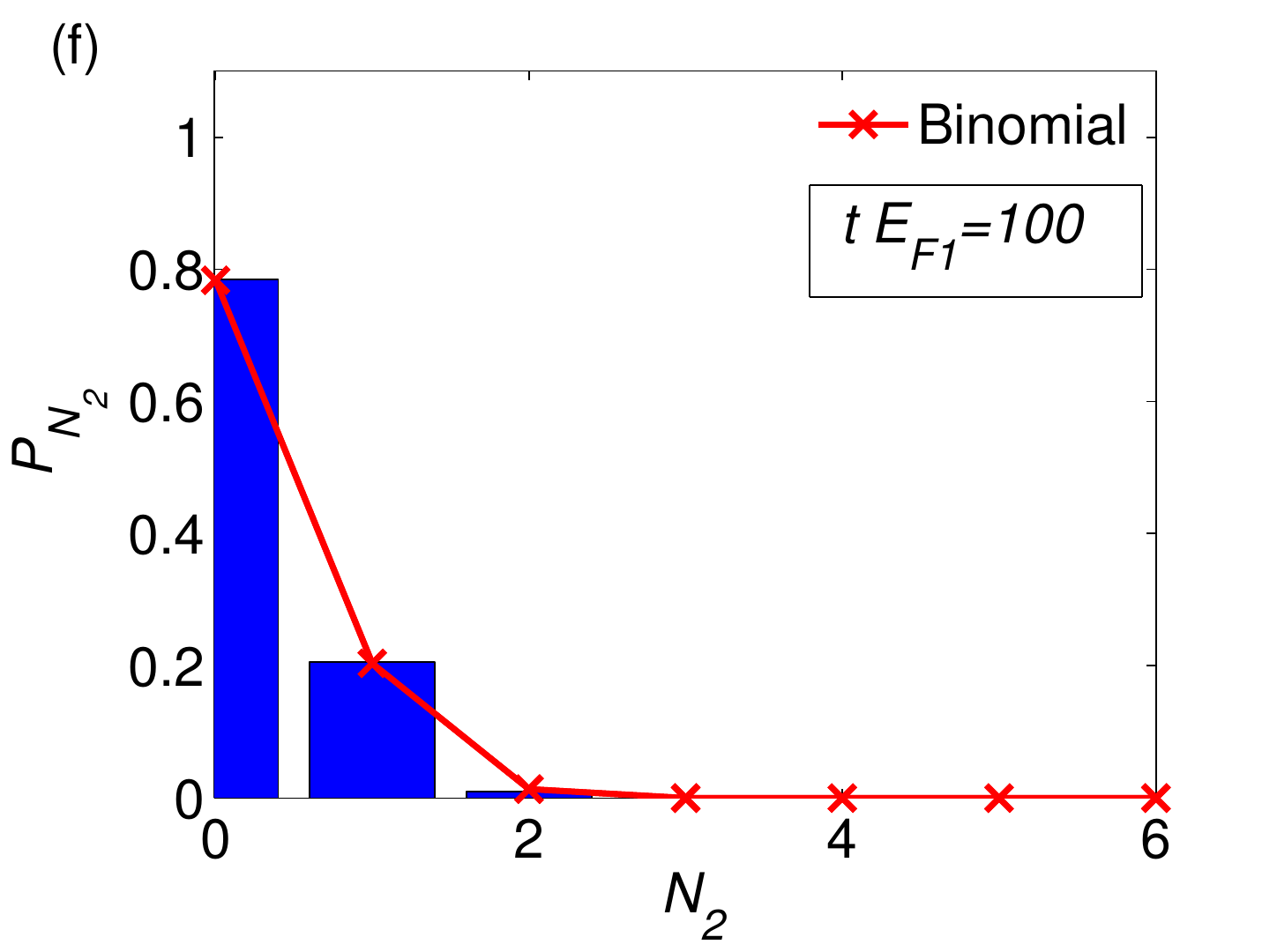}
	\caption{ Non-equilibrium momentum population and FCS obtained in a scenario where in the initial state fermions occupy only a small energy interval. In the  upper panel fermions occupy a low-energy interval  while in the lower panel higher energies are occupied. The second component $|2\rangle$ is initially empty while the  first component $|1\rangle$ has a finite occupation. The interaction strength is  characterized by $k_{F1}a=-0.5$. (a,d) Energy-resolved occupation by the first and second spin component.  The FCS of the number of spin flips $P_{N_2}(t)$ is shown in (b) and (e) for $t E_{F1}=10$, and at $t E_{F1}=100$ in (c) and (f).  The numerical FDA results (blue bars) are compared to a binomial distribution (red crosses).
	}\label{fig_B1}
\end{figure}

To this end, we prepare an initial state where the second Fermi sea of component $\ket{2}$ is empty and where the momentum distribution of the fermions  in the state $\ket{1}$ has only a small interval of energy levels that are occupied. The spin flip rate, as given by Eq.~\eqref{SpinFlipRate}, depends on the scattering phase shift $\delta(E)$ that increases monotonously with $E$ (considering $a<0$). Consequently,  at a fixed interaction strength $k_{F1}a$, fermions in lower energy modes should experience a smaller spin flip rate.

\begin{figure}[t]
	\centering\
	\includegraphics[width=4.6cm]{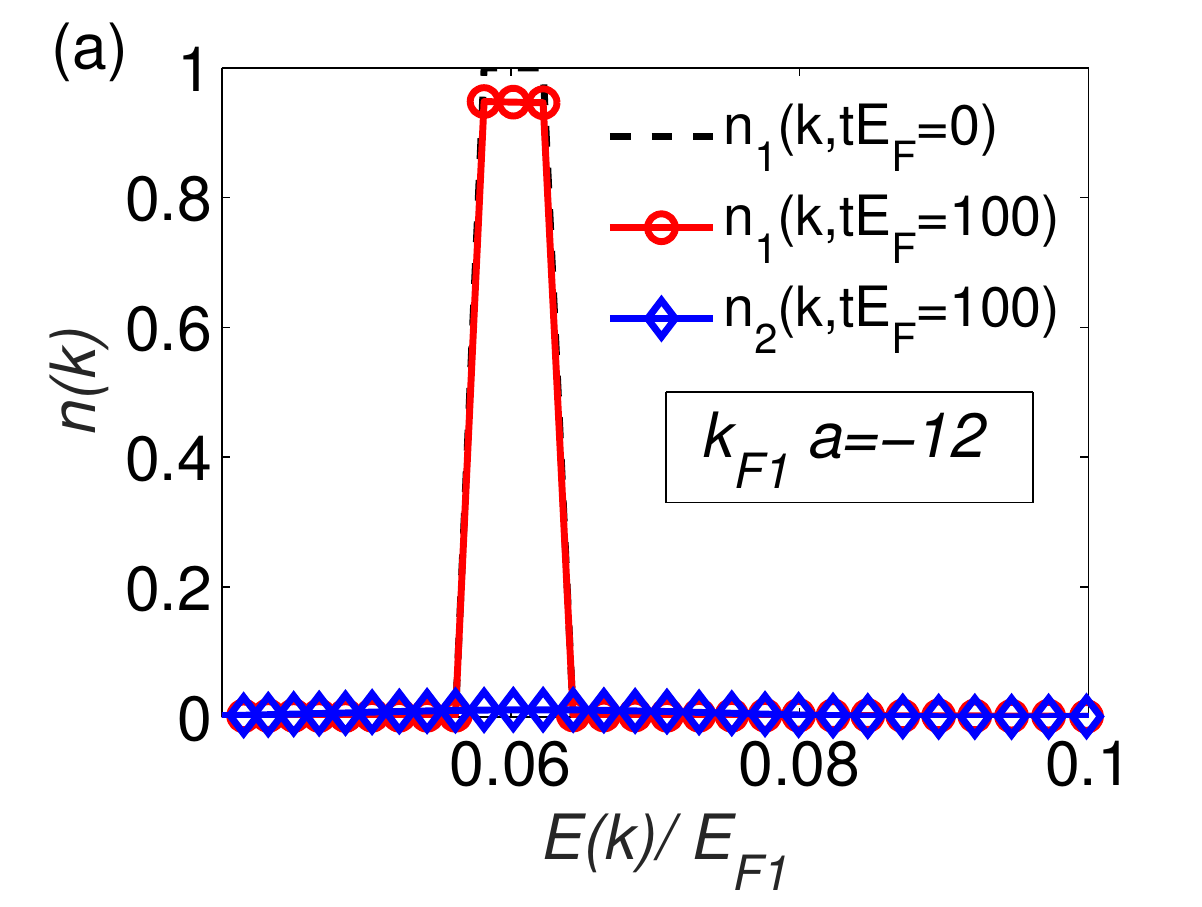}
	\includegraphics[width=4.6cm]{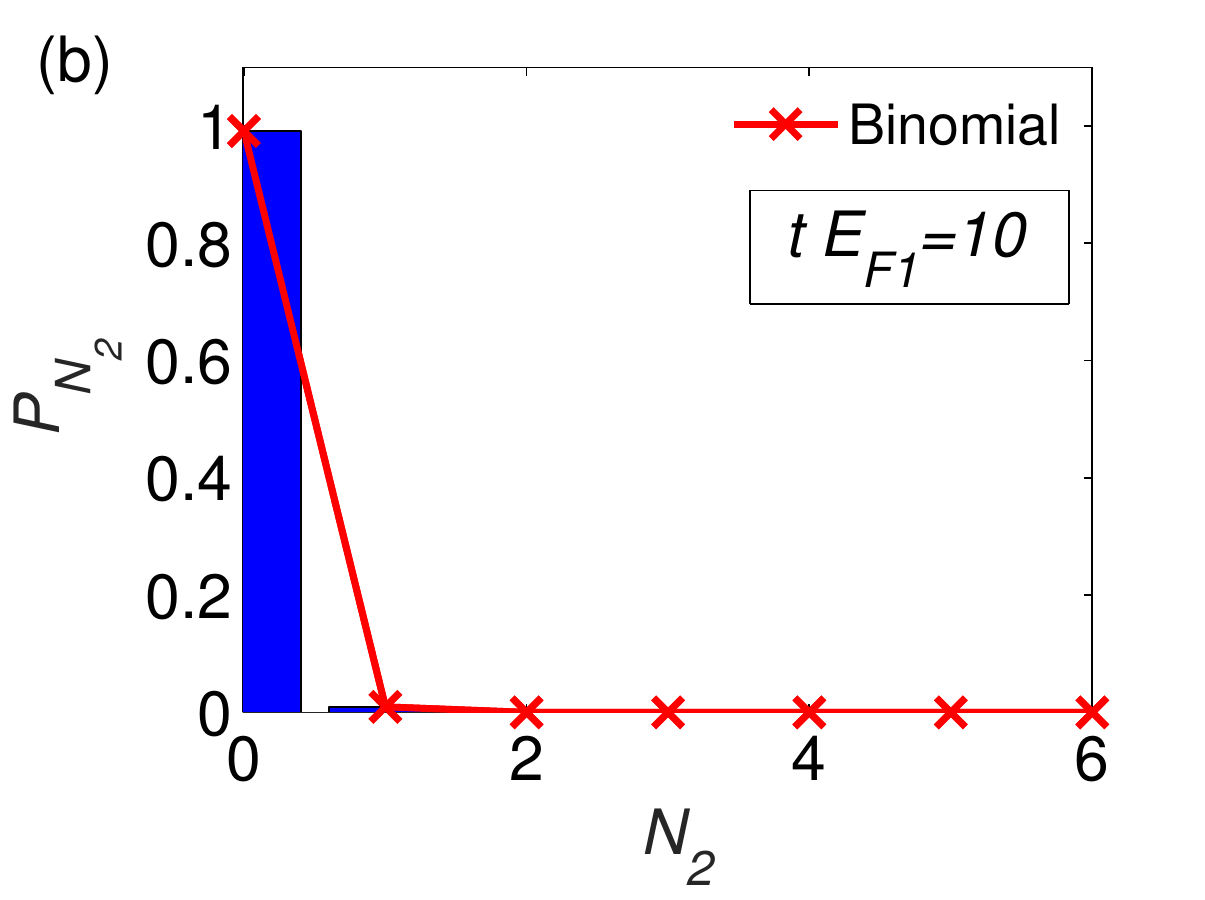}
	\includegraphics[width=4.6cm]{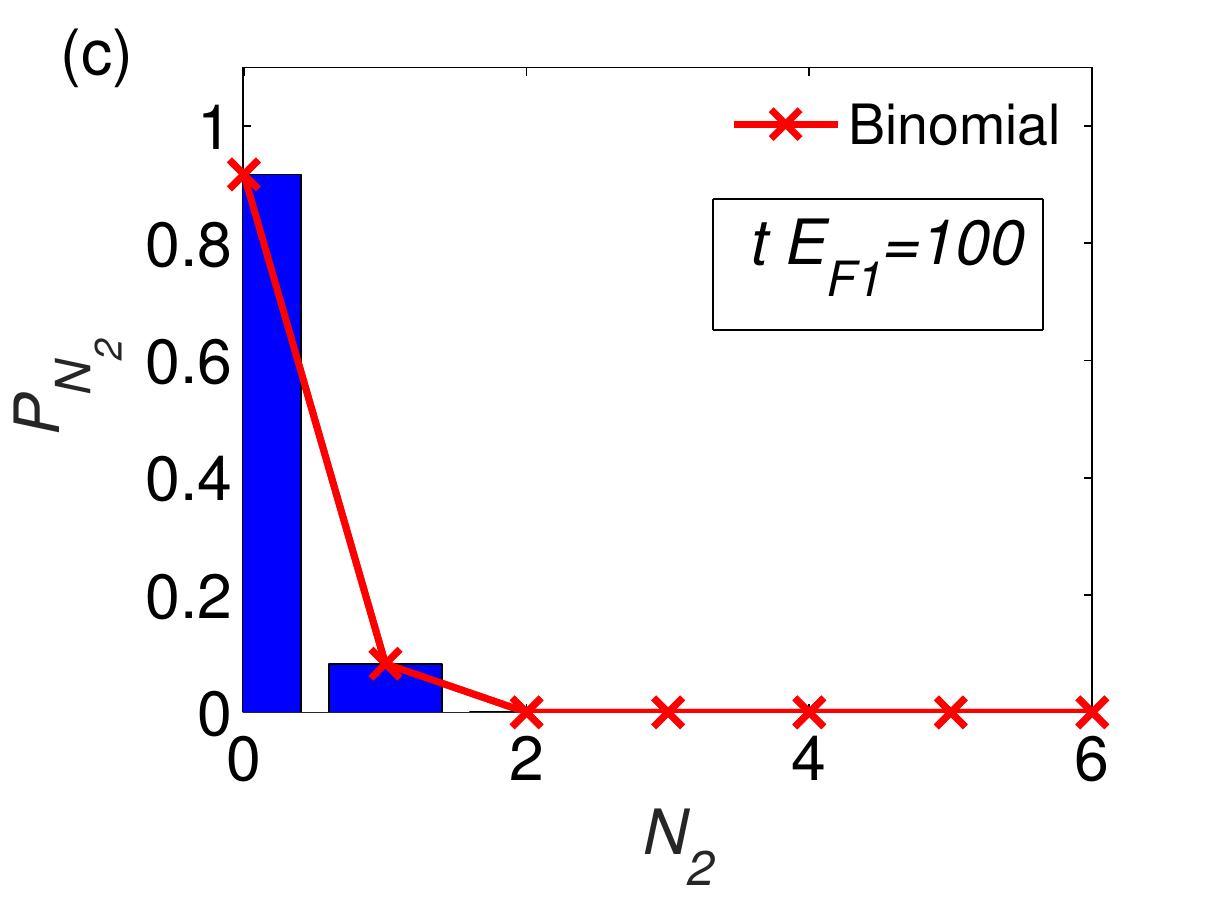}
	\includegraphics[width=4.6cm]{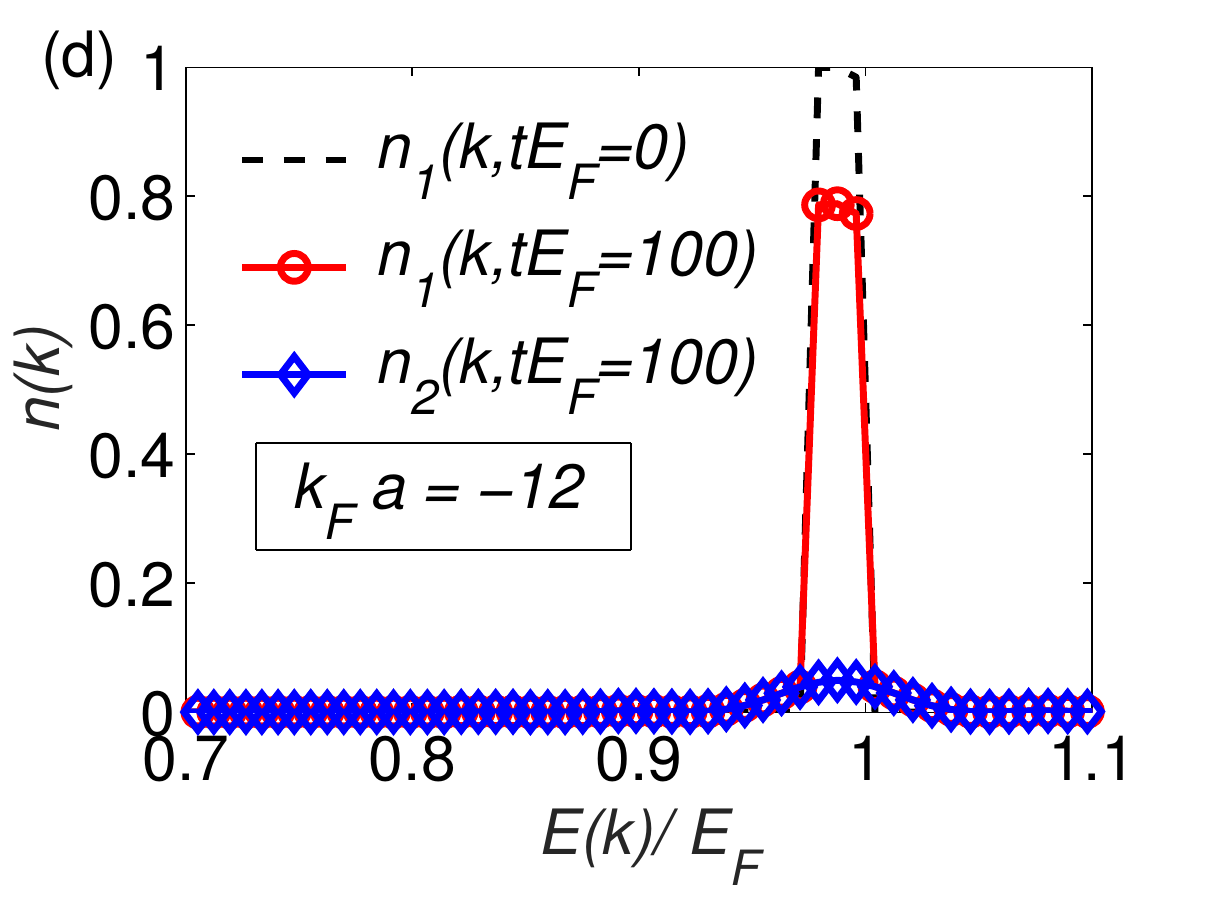}
	\includegraphics[width=4.6cm]{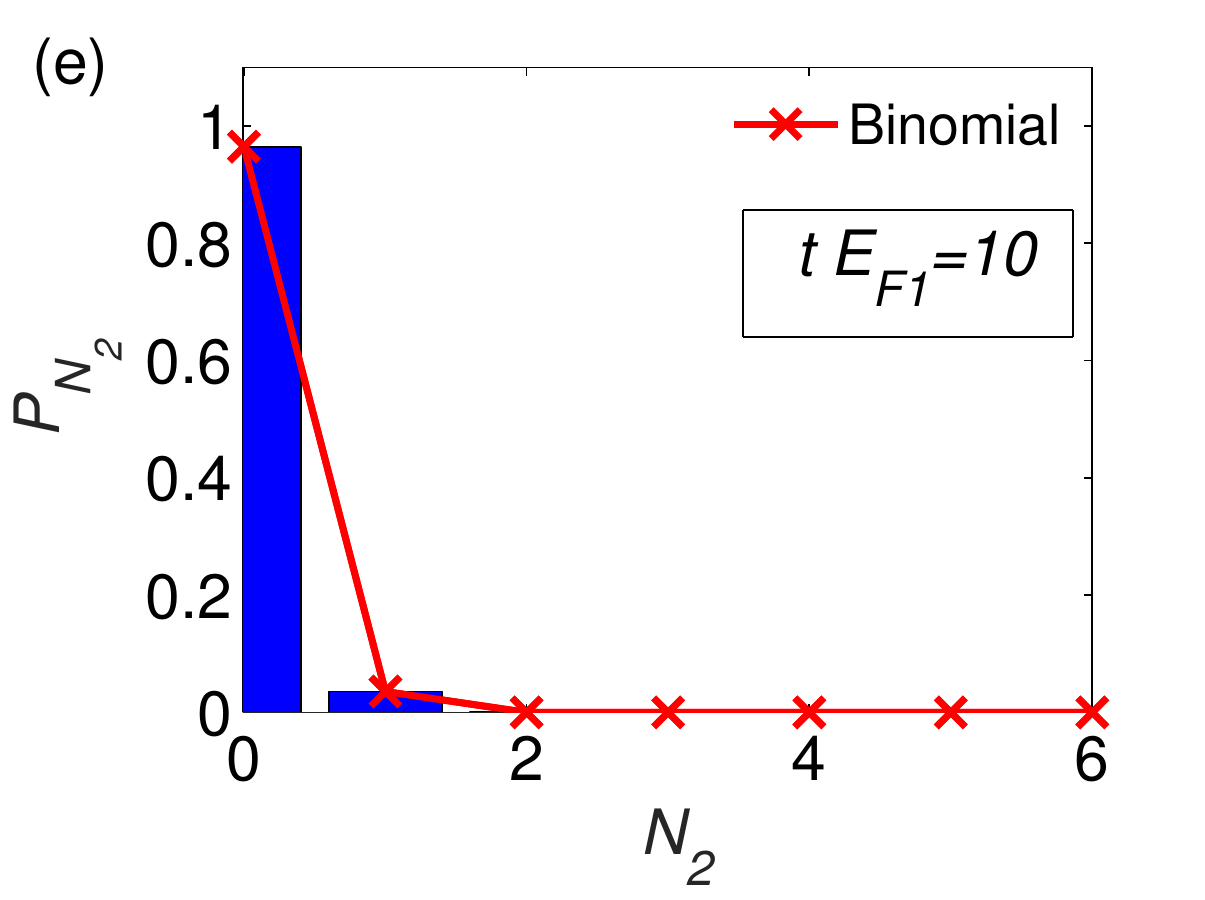}
	\includegraphics[width=4.6cm]{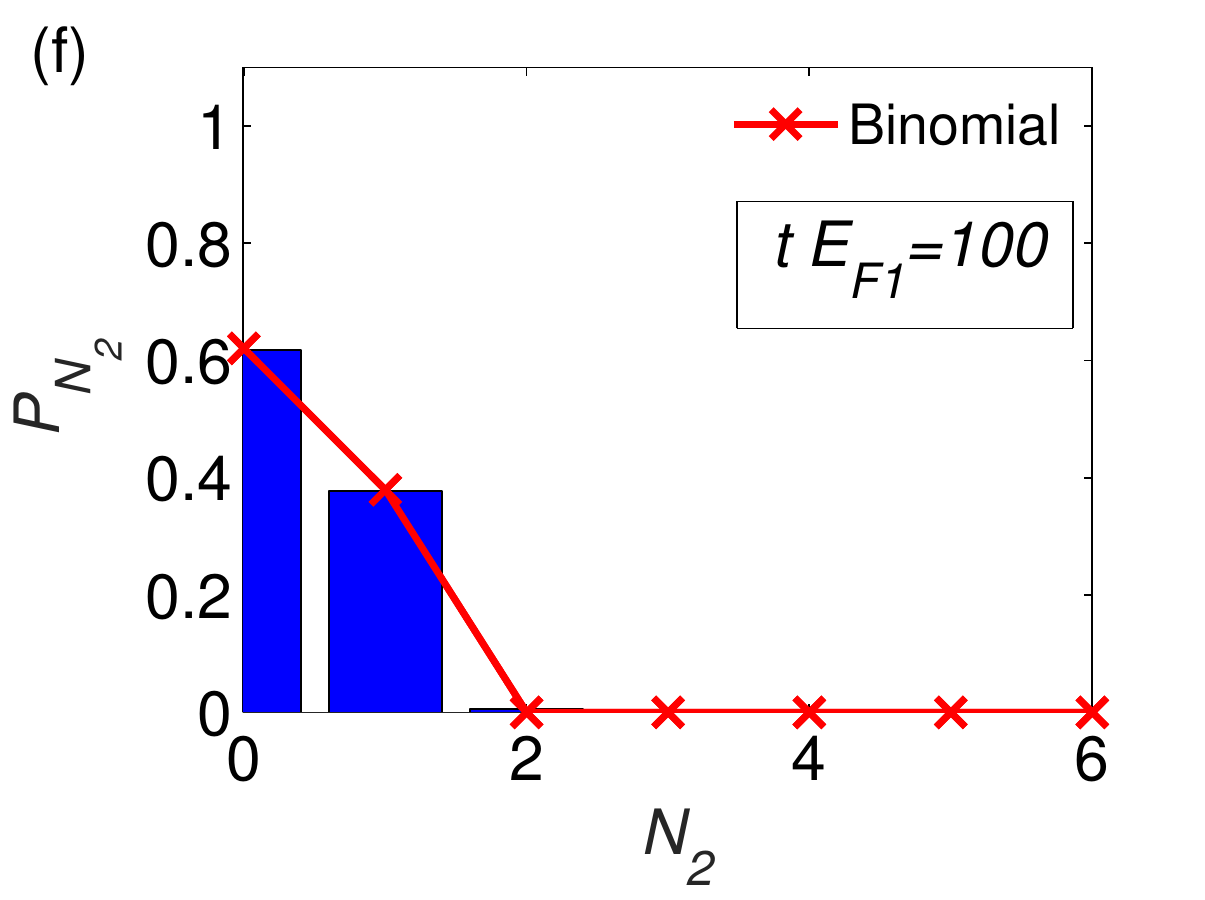}
	\caption{ Non-equilibrium momentum population and FCS starting from initial occupations in a small energy interval as in Fig.~\ref{fig_B1}, here for  interactions characterized by $k_{F1}a=-12$.
	}\label{fig_B2}
\end{figure}

This is confirmed by the numerical simulation shown in Fig.~\ref{fig_B1} for moderate interaction strength $k_{F1}a=-0.5$.  In the upper panel we show the time evolution of the FCS for an initial state occupation confined to a low-energy interval, while for the lower panel higher energy-modes are occupied initially. Confirming our expectation from the analytical result Eq.~\eqref{LevitovChi}, in  both cases the FCS of total spin flips (Fig.~\ref{fig_B1}(b,c) and (e,f), respectively) obeys a binomial distribution. Furthermore, for higher energies the spin flip probability is indeed enhanced. In Fig.~\ref{fig_B2} we repeat the simulation for a  interaction strength $k_{F1}a=-12$ further corroborating our findings. 

Note that in the momentum resolved distributions shown in the left panels of the figures a broadening of the initially sharp distribution function can be seen. This broadening is due to the sudden quench of interactions which projects the initially occupied states into the eigenstates of the interacting Hamiltonian. The overlaps to these states are non-zero also for states outside of the initial energy window which represents the scattering of the fermions to different momentum state upon collisions with the impurity and that  leads  to the broadening of the momentum distribution.

\section{Mapping onto a single-component Fermi gas}\label{sec:appendixC}
The time-dependent response $S(t)$ is obtained from the determinant $\det(\mathbbm{1}+\hat{n}(\hat{R}-\mathbbm{1}))$, where the two-component occupation matrix  $\hat{n}=\textmd{diag}(\hat{n}_1, \hat{n}_2)$ is diagonal in the rotated atomic $(1,2)$-basis. The matrix representing the dynamics, $\hat{R}=\textmd{diag}(e^{i \hat{h}_{0,\uparrow} t/\hbar } e^{-i\hat{ h}_{1,\uparrow} t/\hbar }, \hat{1})=\textmd{diag}(e^{ 2 i \hat{\delta} \theta(t) }, \hat{1})$, is on the other hand diagonal in the non-rotated  basis $(\uparrow,\downarrow)$. In these expressions  $\hat{n}_1/\hat{n}_2$ are
the number operators,  and  $\hat{\delta}$ is the phase shift operator that applies the scattering phase shift to scattering wave packets. 

To compute $S(t),$ we first write both $\hat{n}$ and $\hat{R}$  in the basis $|\uparrow\rangle$ and $|\downarrow \rangle$ using the unitary transformation  $(|1\rangle ,|2\rangle)^T=\hat{U}(|\uparrow\rangle, |\downarrow \rangle)^T,$ with
\be
U=\begin{pmatrix}
	\cos(\frac{\theta}{2}) & -\sin(\frac{\theta}{2})\\
	\sin(\frac{\theta}{2}) & \cos(\frac{\theta}{2})
\end{pmatrix},
\ee
to express $\hat{n}$  as $\hat{n}= \hat{U}^{\dagger}\textmd{diag}(\hat{n}_1, \hat{n}_2)\hat{U}.$
We obtain
\be
S(t)&=&\det\Big[\begin{pmatrix}
	\hat{1} & 0\\
	0 &  \hat{1}
\end{pmatrix}+ \hat{U}^{\dagger} \begin{pmatrix}
\hat{n}_1 & 0\\
0 & \hat{n}_2
\end{pmatrix}  \hat{U} \begin{pmatrix}
e^{ 2 i \hat{\delta} \theta(t) }-\hat{1} & 0\\
0 &  0
\end{pmatrix}
\Big]\\
&=&\det\Big[\begin{pmatrix}
	\hat{1} & 0\\
	0 &  \hat{1}
\end{pmatrix}+   \begin{pmatrix}
\hat{n}_1 \cos^2(\frac{\theta}{2})+ \hat{n}_2 \sin^2(\frac{\theta}{2}) & \frac{(\hat{n}_2-\hat{n}_1)}{2} \sin(\theta)\\
\frac{(\hat{n}_2-\hat{n}_1)}{2} \sin(\theta) & \hat{n}_2 \cos^2(\frac{\theta}{2})+ \hat{n}_1 \sin^2(\frac{\theta}{2})
\end{pmatrix}  \begin{pmatrix}
e^{ 2 i \hat{\delta} \theta(t) }-\hat{1} & 0\\
0 &  0
\end{pmatrix}
\Big]\\
&=&\det(\hat{1}+(e^{ 2 i \hat{\delta} \theta(t) }-\hat{1}) \hat{n}(E))
\ee
where $\hat{n}(E)= \hat n_1 \cos^2 (\theta/2)+\hat n_2 \sin^2 (\theta/2)$ represents a one-component distribution exhibiting  two Fermi surfaces. It is  determined by
\be
{n}(E)=(1-p) n_F(E-E_{F2}) +p n_F(E-E_{F1}),\label{Eq:distribution_B}
\ee
where  we assumed $E_{F2} < E_{F1}$  and defined the polarization angle $p=\cos^2(\theta/2)$.

\section{Fermi surface dynamics from Toeplitz matrices}\label{sec:appendixD}
In this Appendix we study the Fermi surface contributions to the time-dependent overlap function 
\begin{equation}\label{SStart}
S(t)=\langle\psi_F|e^{i \hat{H}_0 t/\hbar } e^{-i \hat{H}_1 t/\hbar }|\psi_F\rangle= \det[1- \hat{n}+\hat{n} e^{i \hat{h}_0 t/\hbar } e^{-i\hat{ h}_1 t/\hbar }]
\end{equation}
using the theory of  Toeplitz  matrices. Here $\hat h_1$ and $\hat h_0$ are the single-particle representations of the many-body Hamiltonian describing the interaction of an impurity with a \textit{single}-component Fermi gas. In Eq.~\eqref{SStart} we have used the mapping onto a single component Fermi gas so that $\hat n$ is the occupation operator given by Eq.~\eqref{Eq:distribution_B}. By inspection of Eq.~\eqref{KlichFormula} it is evident that in this representation the system can be understood to be described by a mixed density matrix. In contrast, without the mapping  the ket $\ket{\psi_F}$ on the LHS of Eq.~\eqref{SStart} represents the pure initial state of the system given by $\ket{\psi_F}=\ket{\text{FS}_1}\otimes\ket{\text{FS}_2}$.

In the following we work in a basis of wave packets localized in time and energy \cite{Hassler2009}. In this basis the time evolution operator $ e^{i \hat{h}_0 t/\hbar } e^{-i\hat{ h}_1 t/\hbar }$ acts approximately diagonally in energy. Following Refs.~\cite{Schmidt2017,Gutman2011}, time may be descretized  according to $t=N\Delta_t$  where we introduce the time interval $\Delta_t=\hbar\pi/ \Lambda$ and a high-energy cutoff $\Lambda$. The overlap $S(t)$ can then be rewritten in terms of an $N\times N$ Toeplitz  matrix $\hat \sigma$:
\be\label{ToeplitzDet1}
S(t)=\det\begin{pmatrix}
	\sigma_0 & \sigma_{-1}& \sigma_{-2}&\cdots & \sigma_{-N+1} \\
	\sigma_1 & \sigma_0& \sigma_{-1}&\ddots & \vdots\\
	\sigma_2 & \sigma_{1}& \ddots &\cdots & \sigma_{-2}\\
	\vdots & \ddots& \ddots &\sigma_0& \sigma_{-1}\\
	\sigma_{N-1} & \cdots& \sigma_{2}&\sigma_1& \sigma_{0}
\end{pmatrix}.
\ee
The matrix elements $\sigma_k$ ($k$ is here a time index) follow from Fourier transformation
\be\label{FTKernel}
\sigma_{k}&=& \int^{2\Lambda}_{0} \frac{d E e^{i E k\Delta_t /\hbar}}{2 \Lambda \hbar} \sigma(E),
\ee
of the kernel
\be\label{regkernel}
\sigma(E)= e^{ i E \delta/\Lambda} (1- {n}(E)+{n}(E) e^{i 2 {\delta}}),
\ee
that is diagonal in energy. The high energy regularization of the integral In Eq.~\eqref{FTKernel} follows from the definition of the time-interval $\Delta_t=\hbar\pi/ \Lambda$ so that energies are restricted to the interval $E\in(0, 2 \Lambda]$. Furthermore,  following Gutman \etal \cite{Gutman2011}, we have imposed a phase factor $e^{i  E  \delta/\Lambda}$ in the kernel $\hat \sigma$ in Eq.~\eqref{regkernel}. 
 Introducing the angular variable $\theta\equiv E \Delta_t/\hbar$  defined  on a unit circle $\theta\in(0, 2 \pi]$ to express $\sigma_{k}=\int^{2 \pi}_{0} \frac{d \theta e^{i \theta k}}{2 \pi} \sigma(\theta)$
 the phase factor  ensures periodicity of the kernel $\sigma(\theta)$ on $(0,2 \pi ]$ in Eq.~\eqref{regkernel}.   In the end of the calculation we will take the limit $\Lambda\to\infty$ so that the phase factor will disappear.  
 
The kernel now obeys  periodic boundary conditions $\lim_{E\rightarrow 0} \sigma(E)= \lim_{E\rightarrow 2\Lambda } \sigma(E)$~\cite{Gutman2011}. This allows us to apply the Szeg\H{o}  theorem \cite{szego1952} to find the asymptotic properties of the Toeplitz  matrix  $S_N$ defined by Eq.~\eqref{ToeplitzDet1} in the limit of large $N$. Since $N=t/\Delta_t$, considering large $N$ corresponds to the  limit of long times $t$. For large $N$ the Szeg\H{o}  theorem states that
\be
\ln \det \hat{\sigma} \sim N [\ln \sigma(\theta)]_0+\sum^{\infty}_{k=1} k [\ln \sigma(\theta)]_k [\ln \sigma(\theta)]_{-k}.\label{eq:Szego}
\ee
The Szeg\H{o}  theorem demands $\ln \sigma(\theta)$ to be be a smooth function with Fourier harmonics $[\ln \sigma(\theta)]_k=\int_0^{2\pi} \frac {d \theta}{2\pi}\ln \sigma(\theta) e^{-i k \theta}$. In our case the smoothness of $\sigma(E)$ is, however, not guaranteed and we rely on the Fisher-Hartwig (FH) conjecture that extends the applicability of Eq.~\eqref{eq:Szego} \cite{Deift2011,Gutman2011,Protopopov2013}. In fact in Ref. \cite{Gutman2011,Protopopov2013} it was shown that  also for  Fermi distributions with multiple steps the naive formula following from the strong Szeg\H{o}  theorem still leads to correct results.

Expressing the real-time overlap function as
\be\label{candEEq}
\det \hat{\sigma}\propto e^{- i \kappa  t + c(t)},
\ee
the first term of Eq.~\eqref{eq:Szego} yields a term with a linear dependence on time,
\be \label{appeq15}
-i t \kappa &=&  \Big\{ N  \int_0^{2\pi} \frac {d \theta}{2\pi}\ln \sigma(\theta) \Big\}\nonumber\\
&=&   \Big\{N  \int^{2\Lambda}_{0} \frac{ \Delta_t d E}{2 \pi \hbar}\ln \sigma(E) \Big\}\nonumber\\
&\rightarrow&  \Big\{ t \int^{\infty}_{0} \frac{  d E}{2 \pi \hbar}\ln (1- {n}(E)+{n}(E) e^{i 2 {\delta}}) \Big\},
\ee
where in the last line we have taken the limit $\Lambda \rightarrow \infty $ so that $\sigma(E)\rightarrow1- {n}(E)+{n}(E) e^{i 2 {\delta}}$.    Accordingly the exponential decay rate $\gamma$, defined by $|S(t)|\sim e^{-\gamma t}$, is given by
\be \label{appeq16}
\gamma&=&  - \textmd{Re} \Big\{ \int^{\infty}_{0} \frac{  d E}{2 \pi \hbar}\ln (1- {n}(E)+{n}(E) e^{i 2 {\delta} }) \Big\}.
\ee

Remarkably, Gutman and coworkers~\cite{Gutman2011} showed that also for subleading contribution $c(t)$ defined in Eq.~\eqref{candEEq} an analytical expression can be found. It is determined by the second term $\sum^{\infty}_{k=1} k [\ln \sigma(\theta)]_k [\ln \sigma(\theta)]_{-k}$ in Eq.~\eqref{eq:Szego} and as shown in \cite{Gutman2011} it leads to a non-trivial power-law behavior in time. Following Ref.~\cite{Gutman2011} we  consider zero temperature $T=0$ and  the double step distribution function ($E_{F2}<E_{F1}$),
\be
{n}(E)=(1-p) \theta(E_{F2}-E) +p \theta(E_{F1}- E).\label{Eq:distribution2}
\ee
The regularized kernel in Eq.~\eqref{regkernel}  takes the form \cite{Gutman2011} 
\be
\sigma(E)=e^{ i E \delta/\Lambda}\times
\left\{
\begin{array}{l@{\;,\;}l}
	e^{2 i   {\delta}} & 0<E< E_{F2}\\
	1+ p( e^{i 2  {\delta}}-1) & E_{F2}<E< E_{F1}\\
	1 & E_{F1}<E\\
\end{array}
\right.
\ee
Thus   $\frac{1}{2 i}\ln\sigma(E)$ can be  expressed as:
\be
\frac{1}{2 i}\ln\sigma(E)=\frac{E \delta}{2\Lambda} +
\left\{
\begin{array}{l@{\;,\;}l}
	\delta & 0<E< E_{F2}\\
	\tilde{\delta}_{\textmd{eff}}(E)\equiv-\frac{i}{2} \ln\big[1+(e^{2 i \delta}-1)p \big]   & E_{F2}<E< E_{F1}\\
	0 & E_{F1}<E\\
\end{array}
\right. .
\ee
The Fourier harmonics $[\ln \sigma(E)]_{k\neq0}=\int_0^{2  \Lambda } \frac {\Delta_t d E}{2\pi\hbar}\ln \sigma(E) e^{-i k \Delta_t E/\hbar}$ required for the evaluation of Eq.~\eqref{eq:Szego} are  given by
\be\label{FTElement}
[\ln \sigma(\theta)]_{k\neq0}=
-\frac{1}{\pi k}[\tilde{ \delta}_1  e^{- i k E_{F1} \Delta_t}+\tilde{ \delta}_2 e^{- i k E_{F2} \Delta_t}]
\ee
where $\tilde{ \delta}_1=\tilde{ \delta}_{\textmd{eff}}(E_{F1}-0^+)$ and $\tilde{ \delta}_2=\delta-\tilde{ \delta}_{\textmd{eff}}(E_{F2}-0^+)$ take into account the phase jumps at  Fermi energies $E_{F1}$ and $E_{F2}$.  From Eq.~\eqref{FTElement} the  second term of Eq.~\eqref{eq:Szego} follows:
\be
\sum^{\infty}_{k=1} k [\ln \sigma(\theta)]_k [\ln \sigma(\theta)]_{-k}&=&-\sum^{\infty}_{k=1}  \frac{1}{\pi^2 k }[\tilde{ \delta}_1 ^2+\tilde{ \delta}_2^2+2 \tilde{ \delta}_1 \tilde{ \delta}_2 \cos\big(k \Delta_t (E_{F1}-E_{F2})\big)]\nonumber\\
&\sim& -\int^t_{\Delta_t} d\tau  \frac{1}{\pi^2 \tau }[\tilde{ \delta}_1 ^2+\tilde{ \delta}_2^2+2 \tilde{ \delta}_1 \tilde{ \delta}_2 \cos\big(\tau (E_{F1}-E_{F2})\big)].
\label{eq:secondterm}
\ee
When $E_{F1}=E_{F2},$ Eq.~\eqref{eq:secondterm} gives $-\frac{\delta^2}{\pi^2}\ln \frac{t \Lambda}{\pi},$ which recovers correctly the power-law decay of $S(t)$ characteristic for the Anderson OC that considers a single-component  Fermi sea with a single Fermi edge. For $E_{F1}\neq E_{F2}$ in the long-time limit defined by $t |E_{F1}-E_{F2}| \gg 1$,  Eq.~\eqref{eq:secondterm} leads to
\be \label{secondterm}
\sum^{\infty}_{k=1} k [\ln \sigma(\theta)]_k [\ln \sigma(\theta)]_{-k}&\sim& -( \frac{\tilde{ \delta}_1^2 }{\pi^2}+\frac{\tilde{ \delta}_2^2}{\pi^2}) \ln \frac{t \Lambda}{\pi}-2 \tilde{ \delta}_1 \tilde{ \delta}_2 [\ln \frac{\pi(E_{F1}-E_{F2})}{\Lambda}+\gamma-\frac{\pi^2(E_{F1}-E_{F2})^2}{4\Lambda^2}]
\ee
where we applied the limit  $\Delta_t\rightarrow0$ (i.e. $\Lambda\to\infty$) and performed the cosine integral $C_i(x)=\int^{\infty}_x du \frac{\cos(u)}{u}\approx\ln x+\gamma-\frac{x^2}{4}$ with $\gamma$ the Euler-Mascheroni constant.  Note that the last term in Eq.~\eqref{secondterm}, which is proportional to  $\tilde{ \delta}_1 \tilde{ \delta}_2$, is time-independent.

Combining all results for the case of two Fermi steps assuming $E_{F2}<E_{F1}$ we obtain the long-time behavior of the contribution from particle-hole excitations at the two Fermi edges as
\be
S^{(FS)}(t)\propto t^{-(\frac{\tilde{ \delta}_1^2 }{\pi^2}+\frac{\tilde{ \delta}_2^2}{\pi^2})} e^{- i  \kappa_0 t }.
\ee
Thus we identify  $S_0^{(FS1)}(t)\sim t^{-\frac{\tilde{ \delta}_1^2 }{\pi^2}}$ at Fermi edge $E_{F1}$ and $S_0^{(FS2)}(t)=t^{-\frac{\tilde{ \delta}_2^2}{\pi^2}}$ at the Fermi edge $E_{F2}$ where
\be
\tilde{\delta}_1&=&\tilde{\delta}_{\textmd{eff}}(E_{F1}-0^+)\\
\tilde{\delta}_2&=&\tilde{\delta}_k(E_{F2}+0^+)-\tilde{\delta}_{\textmd{eff}}(E_{F2}-0^+).
\ee
Inspired by previous studies of Fermi surface contributions with $n\neq 0$ for the case of an impurity interacting with a single-component Fermi gas in its ground state \cite{Schmidt2017} we may now straightforwardly conjecture the generalization to our case of a spin-flip Hamiltonian \eqref{SpinPumpHamiltonian} and arrive at
\begin{align}\label{bbformula2}
S_n^{(FS1)}(t)&\propto e^{- i n E_{F1} t}  \Big(\frac{1}{ t}\Big)^{(\frac{\tilde{\delta}_1}{ \pi}-n)^2},\nonumber\\
S_n^{(FS2)}(t)&\propto e^{- i n E_{F2} t} \Big( \frac{1}{ t}\Big)^{(\frac{\tilde{\delta}_2}{ \pi}-n)^2}.
\end{align}

Finally we note, that in Eqs. \eqref{appeq15} and \eqref{appeq16}  one may  reintroduced the energy dependent phase shift ${\delta}(E)$ on a phenomenological basis and also apply those results to the case of finite temperature. Indeed we find that these expressions  yield excellent agreement with exact numerical results for a large range of temperatures (see Fig.~\ref{fig_7}).  In fact Eqs. \eqref{appeq15} and \eqref{appeq16} represent a direct generalization of previous findings \cite{Schmidt2017} which were restricted to the case of an impurity interacting with  a Fermi gas with a single Fermi-step distribution $n(E)$, to the case of  non-equilibrium fermions with a multi-step distribution that fulfills $n(E)=1$ for $E=0$ and $n(E)=0$ for $E\rightarrow\infty.$

\section{Relation of Ramsey decoherence and FCS}\label{app.ramsey.FCS}

The exponential decay rate of the Ramsey signal at long times at $T=0$ and $E_{F2}=0$ is determined by
\begin{equation}
A=-\gamma=\ln |S(t)| = \text{Re} \ln S(t) = t \int^{E_{F1}}_0 \frac{d E}{2 \pi} \text{Re}\ln[ 1+p( e^{2 i {\delta}(E)}-1)]\label{eqapp1}
\end{equation}
where $p=\cos^2\theta/2$ is the polarization angle. Using $\text{Re}\ln z= \ln|z|$ one finds
\begin{align}
A&= t \int^{E_{F1}}_0 \frac{d E}{2 \pi} \ln\left| 1+p( e^{2 i {\delta}(E)}-1)\right|\nonumber\\
&= t \int^{E_{F1}}_0 \frac{d E}{2 \pi} \frac{1}{2}\ln\left[ 1-2p(1-p)( 1-\cos 2\delta(E))\right].\label{appequiv}
\end{align}
Now consider the quantity
\begin{align}
B= \frac{1}{2}\ln\chi(e^{i \lambda}\to 0)
\end{align}
where $\chi$ is given by Eq.~\eqref{LevitovChi}, so that
\begin{align}
\ln \chi (e^{i \lambda}\to 0)=t\int_0^{E_{F1}} \frac{dE}{2\pi\hbar} \ln[1-\Gamma(E)].
\end{align}
Using $\Gamma(E)= \sin^2\theta \sin^2\delta(E)$, $\sin^2\theta = 4 p (1-p)$ and $\sin^2 \delta(E) = (1-\cos 2\delta)/2$ one finds that indeed
\begin{equation}
B= t \int^{E_{F1}}_0 \frac{d E}{2 \pi} \frac{1}{2}\ln\left[ 1-2p(1-p)( 1-\cos 2\delta(E))\right]
\end{equation}
which equals Eq.~\eqref{appequiv} and hence we have shown
\begin{equation}
|S(t)|\to \sqrt{\chi(e^{i\lambda}\to0 )}. 
\end{equation}
This prescription  projects out the contribution $N_2=0$ in Eq.~\eqref{expansion}, so that we can indeed conclude that up to logarithmic corrections,
\begin{equation}
|S(t)|= \sqrt{P_{N=0}(t)} = \left[\int d \lambda \chi(\lambda,t)\right]^{1/2}.
\end{equation}

\section{FCS for  a finite number of  impurities}\label{app_CLT}

Experiments that use  impurities as  probes, are naturally subject to relatively small signal-to-noise ratios due to the small numbers of impurities.  By using the many-body medium itself as a probe, our experimental scheme circumvents this challenge. In particular, the measured signal can become large at late times, because the impurity can flip an arbitrary number of spins in the background gas.  The fact that many spin flips occur has also a  consequence for theoretical approaches to the impurity-induced spin-transport problem: Since the number of spin-flipped atoms easily exceeds one, simple variational wave functions based on few-fermion excitations \cite{Chevy2006,Punk2009,Schmidt2012b,Massignan2014,Schmidt2015ang,Sidler2016,lemeshko2018,liu2018} are bound to  fail. 
 
In typical experimental setups the impurity number will be finite which rises the question of what the influence of a finite density of impurities is on the observed dynamics. In this regard the typical inter-particle distance $d\sim n_I^{-1/3}$ between impurities of a density $n_I$ becomes a relevant length scale. As a very conservative estimate, the dynamics will be governed by the physics of independent scattering centers as long as times $t v_F<d$ (with $v_F$ the Fermi velocity) are considered. Only when $t v_F>d$ fermions will be able to scatter from multiple impurities leading to correlated scattering events that are, for instance,  the basis for  bath-mediated, Ruderman-Kittel-Kasuya-Yosida (RKKY)-type, impurity-impurity interactions.

Here, we focus on the regime of a low-impurity density where induced interactions can be neglected. In this case scattering events are independent and each impurity (representing an independent stochastic variable) is characterized by a FCS with generating function $\chi(\lambda)$. The probability $\bar P_{N_2^\text{Tot}}$ to measure a total number $N_2^\text{Tot}$ of spin flipped atoms  in a sample of $N_I$ impurities (localized in a central region of a Fermi gas of constant density)  is then derived from the characteristic function
\begin{equation}\label{FCSTot}
\chi^\text{Tot}(\lambda,t) = \left[\chi(\lambda,t)\right]^{N_I}.
\end{equation}
The evaluation of the Fourier transform of this expression yields the desired probability 
\begin{equation}
\bar P_{N_2^\text{Tot}}(t) = \int d\lambda \left[\sum_{N_2}P_{N_2}(t) e^{i \lambda N_2}\right]^{N_I} e^{-i \lambda N_2^\text{Tot}}.
\end{equation}
This equation renders the constraint $N_2^\text{Tot} = \sum_i N_2(i)$, where $N_2(i)$ is the number of spin flips produced by the $i$-th impurity, particularly transparent. As we have seen, the distribution $P_{N_2}$ is well described by a sum over binomials, c.f. Eq.~\eqref{LevitovChi}, so that $P_{N_2}$ has well-defined moments. Thus, by virtue of the central limit theorem,  the distribution of total observed spin flips, $\bar P_{N_2^\text{Tot}}$, approaches  a normal distribution for a sufficiently large number of impurities  $N_I$.

\begin{figure*}[tb]
\centering
\includegraphics[width=1\textwidth]{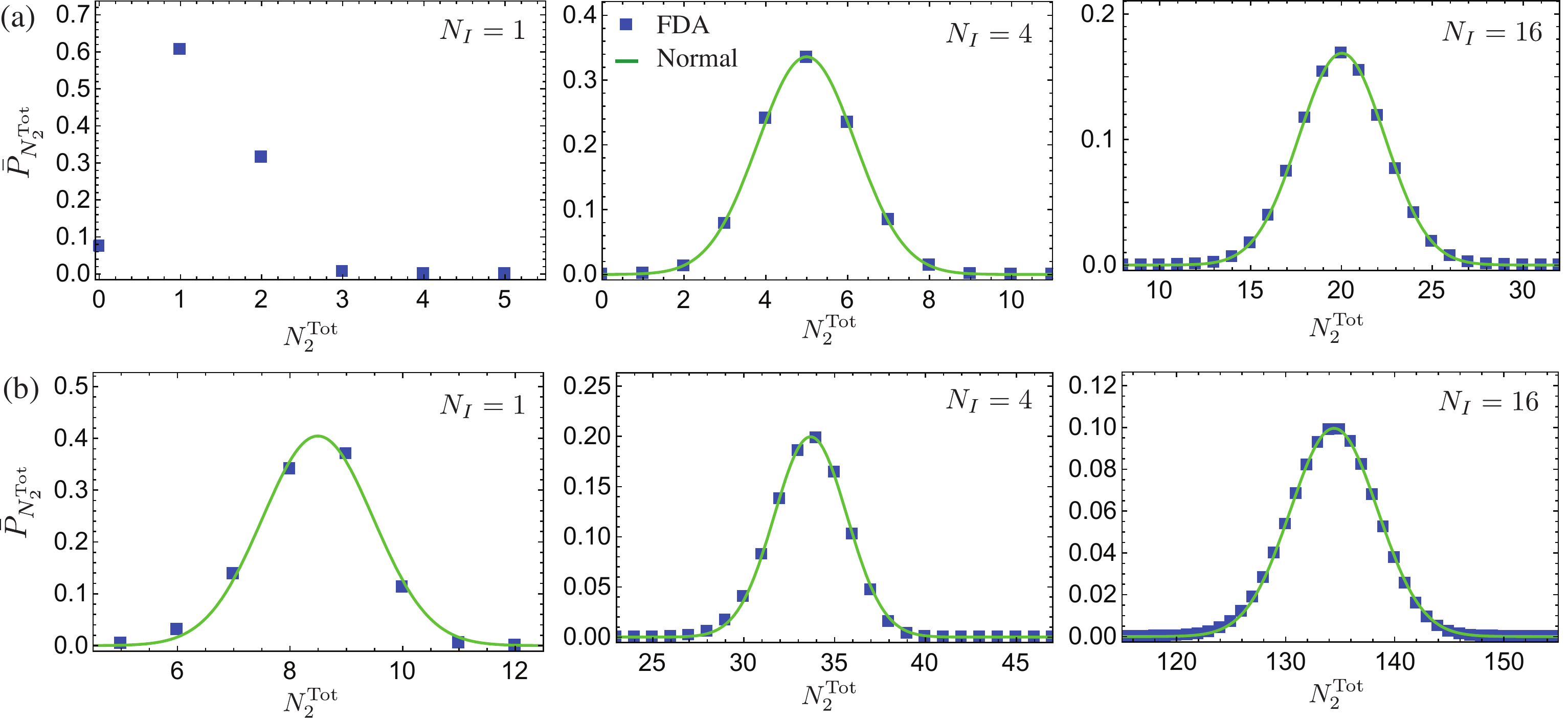}
\caption{ \textbf{Influence of multiple impurities.} FCS $\bar P_{N_2^\text{Tot}}$ of total number of spin flips $N_2^\text{Tot}$  for strong interactions $k_{F1}a=-6$ at times (a) $tE_{F1}=10$  and (b) $tE_{F1}=60$ for $N_I=1,4,16$ impurities immersed in the Fermi gas (left to right). The blue squares represent the exact result from FDA. For the first figure in (a) the normalized Gaussians is not shown. It does not fit the data since $N_2^\text{Tot}$ is bound by zero from below.
}\label{fig_CFL}
\end{figure*}

This can be seen explicitly as follows:  let us assume that the impurities represent independent and identically distributed random variables  ${\hat{N}(1), \cdots, \hat{N}(N_I)}$, each with mean value $ \langle \hat{N} \rangle$ and variance $\sigma_N^2$. Consequently  $\sum^{N_I}_{x=1}\hat{N}(x)$ has mean vale $N_I \cdot\langle \hat{N} \rangle$ and variance $N_I\cdot \sigma_N^2$. As dictated by the central limit theorem, the probability $P_{N_2^\text{Tot}}$ will tend towards a normal distribution as  the number of independent random variables increases. To make this statement more precise we follow standard textbooks \cite{Lemons2002}, and define the   sum of rescaled variables
\be
\hat{Z}_{N_I}=\sum^{N_I}_{x=1}\frac{1}{\sqrt{N_I}}\hat{Y}_x
\ee
where the variables $\hat{Y}_x=\frac{\hat{N}(x)-\langle \hat{N} \rangle}{\sigma_N}$ have zero mean and unit variance.
The characteristic function of $Z_{N_I}$ is
\be
\chi_{\hat{Z}_{N_I}}(\lambda)=\chi_{\sum^{N_I}_{x=1}\frac{1}{\sqrt{N_I}}\hat{Y}_x}(\lambda)=\chi_{\hat{Y}_1}(\frac{\lambda}{\sqrt{N_I}})\chi_{\hat{Y}_2}(\frac{\lambda}{\sqrt{N_I}})\cdots\chi_{\hat{Y}_{N_I}}(\frac{\lambda}{\sqrt{N_I}})=\big[\chi_{\hat{Y}_1}(\frac{\lambda}{\sqrt{N_I}}) \big]^{N_I}
\ee
where we made use of the fact that $\chi_{\frac{ \hat{Y}_x}{ \sqrt{N_I}  } }(\lambda)=\chi_{\hat{Y}_x}(\frac{\lambda}{\sqrt{N_I}  }).$
By expanding the characteristic function $\chi_{\hat{Y}_1}(\frac{\lambda}{\sqrt{N_I}})$,
\be
\chi_{\hat{Y}_1}\left(\frac{\lambda}{\sqrt{N_I}}\right) &=& \sum_{ N_1} P_{ N_1} e^{i \frac{\lambda}{\sqrt{N_I}}  \frac{N_1-\langle \hat{N} \rangle}{\sigma_N}}=  1+\frac{i ^2\lambda^2}{2 N_I}  +O\left(\left(\frac{\lambda}{\sqrt{N_I}}\right)^3\right)
\ee
 the characteristic function   $\chi_{\hat{Z}_{N_I}}(\lambda)$ can be written as
\be
\chi_{\hat{Z}_{N_I}}(\lambda)\simeq\Big( 1+\frac{i ^2\lambda^2}{2 N_I} \Big)^{N_I}\rightarrow e^{-\frac{\lambda^2}{2}}
\ee
where we have used $e^x=\lim_{n\rightarrow 0}(1+x/n)^n$. This last expression shows  that, even when the probability distribution of  a single impurity, obtained from  $\langle e^{i \lambda  \hat{N}}\rangle$, is not Gaussian, the  distribution of the  $\sum^{N_I}_{x=1}\hat{N}(x)$ indeed becomes a normal distribution  as $N_I \rightarrow \infty$, in accordance with  the central limit theorem.

In Fig.~\ref{fig_CFL}  we show the spin flip dynamics at strong interactions $k_{F1}a = -6$, for up to $N_I=16$ impurities immersed in a Fermi gas. At long times and for such strong interactions a normal distribution is quickly approached. In this figure we assume that the spatial inter-impurity separation is chosen such that up to the maximal times shown, $tE_{F1}=60$, scattering events can be treated as independent. As discussed above,  beyond this time scale, multi-impurity collisions will affect the normal distribution at late times in a non-trivial way, which would be intriguing to measure experimentally.

\end{document}